\documentclass[12pt]{article}

\usepackage{amssymb,amsmath,amsfonts,authblk,booktabs,blindtext, eurosym,geometry,ulem,graphicx,caption,color,setspace,sectsty,comment,footmisc,caption,pdflscape,subfigure,array,hyperref,float, multirow, pbox, rotating, adjustbox, caption, regexpatch, siunitx, setspace,tabularx}
\usepackage[english]{babel}
\usepackage[flushleft]{threeparttable}
\usepackage{array}
\usepackage{ragged2e}
\usepackage{multicol}
\usepackage{verbatim} 
\usepackage{graphicx}
\usepackage{adjustbox}
\usepackage{tabularx}
\usepackage{xurl}
\usepackage[colorinlistoftodos]{todonotes}
\usepackage{atbegshi}% http://ctan.org/pkg/atbegshi
\usepackage{booktabs}
\usepackage{threeparttable}
\usepackage[round]{natbib}
\usepackage{authblk}

\definecolor{darkblue}{rgb}{0.0,0.0,0.5}

\newcolumntype{P}[1]{>{\RaggedRight\hspace{0pt}}p{#1}}

\hypersetup{
	colorlinks= true,
	linkcolor = darkblue,
	urlcolor  = darkblue,
	citecolor = darkblue,
	anchorcolor = blue
}
\begin{document}
\title{Transaction Costs and Speed in the Ethereum Ecosystem: Scalability of the Mainnet and Layer 2s.}
\date{June 2026}
\author[1]{Meghan Ambrosia and \qquad Bruce Mizrach\thanks{Correspondence: Department of Economics, Rutgers University, 75 Hamilton Street, New Brunswick, NJ 08901 USA. email: mizrach@econ.rutgers.edu.}}
\affil[1]{\small Department of Economics, Rutgers University, New Brunswick, NJ USA}

\begin{titlepage}
\maketitle

\begin{abstract}
We study the evolution of transaction speed and fees from January 2024 through March 2026, comparing Ethereum Mainnet and its Layer 2 (L2) networks, as well as Solana and Polygon. Ethereum has undergone upgrades that have increased block size and blob count. These upgrades have doubled transactions per second (TPS) on both the Mainnet and the L2 networks. Mainnet median fees have fallen from over \$2 to under \$0.02, and L2 median fees have fallen more than 95\% from \$0.05 to \$0.0015. We forecast that Mainnet median fees will converge with Solana's in August 2027, but TPS will remain below 100 until 2034. The L1 Strawmap, proposing EIP-7938, a potential exponential increase in the gas limit, brings the Mainnet to only 100 TPS in January 2028.  With continued blob expansion, L2s will surpass Solana's TPS in March 2029 and have lower median fees by October 2026.

\end{abstract}

\setcounter{page}{0}
\thispagestyle{empty}
\end{titlepage}
\pagebreak \newpage

\section{Introduction}
Transaction speed and cost are defining competitive dimensions of all financial markets. Equity markets have long set the standard, offering free commissions and millisecond execution for retail orders.\footnote{\url{https://www.fidelity.com/trading/execution-quality/overview}} Recent Ethereum blockchain upgrades are enabling the Mainnet and its adjacent Layer 2 networks (L2s) to close the gap with traditional financial markets and the fastest blockchains, most notably Solana. 

The two key improvements to the Mainnet and its L2s are increased gas limits and the introduction of blobs. A \textit{gas limit} is defined as the maximum amount of computational work (\textit{gas}) permitted into a block. The number of transactions that can be contained in a block on the Mainnet is constrained by the gas limit value. Binary Large Objects (\textit{blobs}), introduced in the Dencun upgrade in March 2024, are the mechanism by which L2s post their compressed transaction data to the Mainnet. 

We compare transaction speeds and fees across Ethereum and its L2s with Solana and Polygon, from January 2024 through March 2026. This window captures the impact of the three most recent Ethereum protocol upgrades: Dencun in March 2024, Pectra in May 2025, and Fusaka in December 2025. Two blob-parameter-only (BPO) upgrades, which increase the number of blobs permitted per block, are activated in December 2025 and January 2026. The period also corresponds with significant increases to the block gas limit, rising from a median daily gas limit of 30 million in January 2024 to 60 million by December 2025.  

Daily average Mainnet transactions per second (TPS) rose from 13.67 in the first quarter of 2024 (2024 Q1) to 25.78 in the first quarter of 2026 (2026 Q1). L2 totals rose from 78.60 to 226.92 TPS over the same time period. Solana TPS did not grow as quickly, rising from 804.72 in 2024 Q1 to 1,303.46 in 2026 Q1; Polygon rose slightly faster than the Mainnet, rising from 45.72 TPS in 2024 Q1 to 91.51 TPS in 2026 Q1.

Mainnet speeds are increasing with the block gas limit.  We estimate that a ten-million-unit increase in the gas limit is
associated with 3.52 more TPS. The Mainnet, even with the gas limit at more than four times the current 60 million, is not projected to reach 100 TPS until February 2034, less than one-tenth of Solana's current speed.

There are ongoing proposals to increase the gas limit as outlined in the L1 Strawmap.\footnote{\url{https://strawmap.org}} The Ethereum Improvement Proposal (EIP)-7938\footnote{\cite{eip7938}} presents a hundred-fold gas limit increase for four years beginning in 2027. Applying this to our current Mainnet projection, we predict the Mainnet to reach 100 TPS by January 2028, six years earlier than our baseline trend.

We find that blob parameter increases on the Mainnet have increased the speed of the L2s.  We estimate that each additional target blob is associated with 10.75 additional L2 total TPS. With continued blob expansion at the BPO increment rate, L2 total TPS is projected to surpass Solana's predicted TPS by March 2029, converging at 1,820 TPS. 

Our manuscript then turns to transaction costs. Transaction fees on the Mainnet have fallen substantially, with a median transaction fee of \$3.786300 in 2024 Q1, declining 99.68\% to \$0.012283 in 2026 Q1. The average fees of the three highest volume L2s: Arbitrum, Base, and Optimism, which we call the L2 leaders, declined 99.16\% from \$0.180219 in 2024 Q1 to \$0.001512 in 2026Q1. While Solana's fees also fell, the percentage decrease was slower. Median fees fell from \$0.000687 in 2024 Q1 to \$0.000496, a decline of  27.80\%. Polygon fees are an exception to this trend, rising 19.62\% over this period from \$0.005742 to \$0.006868. 

As with speed, we find that gas limits are a key factor in declining transaction costs.  We estimate that every five-million-unit increase in the gas limit is associated with a 59.2\% decline in median transaction fees.  The Mainnet is projected to converge with Solana's predicted median transaction fee in August 2027.

For the L2 leaders, each additional target blob is associated with a 4.3\% decline in median transaction fees.  Assuming continued blob expansion at the BPO increment rate, we project that the L2 transaction fees will converge with Solana's predicted fee in October 2027.

The remainder of our paper is organized as follows. Section \ref{sec:lit} reviews related literature, Section \ref{sec:blobsandgaslimits} describes the Ethereum network architecture and documents the sequence of recent protocol upgrades. Sections \ref{sec:pol} through \ref{sec:l2n} cover Polygon, Solana, and the L2s. Section \ref{sec:blockchain_compare} reports transaction speed and fee comparisons across blockchains. Sections \ref{sec:tps_forecast} and \ref{sec:feeforecast} estimate the effects of gas limits and blobs on speed and transaction costs. From these models, we produce forecasts for both the Mainnet and the Layer 2s for TPS and fees.  Section \ref{sec:con} concludes.

\section{Literature}
\label{sec:lit}
This literature covers previous empirical work on Ethereum transaction dynamics and blockchain comparisons.

\subsection{Transaction Speed on Ethereum}

There is established literature covering Ethereum transaction speeds, and there are different methods of defining transaction time. \citet{pow} finds no relationship between the amount of gas and gas price and the delay time in confirming transactions.  Studying EIP-1559,\footnote{\cite{eip1559}} which introduced the base fee, \citet{LiuEIP1559} finds that the EIP reduced the time between when a transaction enters a mempool and when it is mined. Additionally, they find that when Ether's price is volatile, wait times increase.  Conversely, \citet{pacheco2023ethereum} finds that recent gas prices are the most important feature in explaining block inclusion. Our paper differs from this literature in that we measure transaction speed after block formation.

\citet{nethermind2025gasbenchmark} examines the ability of various clients, including Nethermind, to support higher gas limits and concludes that expanded block space increases TPS.   \citet{Ta2024} find that raising the block gas limit increases the number of transactions per block even at shorter block formation times.

\subsection{Transaction Costs on Ethereum}
Studying fees from November 2017 through January 2019, \citet{DonmezKaraivanov2022} finds that the effect of network congestion on Ethereum transaction fees is statistically insignificant until blocks reach 90\% capacity. On the supply side, they find that increasing the block's capacity limit is associated with lower fees. Studying the Ethereum network's transition from Proof-of-Work (POW) to its current transaction method, Proof-of-Stake (POS), \citet{JAINBlockchainFees2023} finds that Ethereum fees in USD increase nonlinearly with mempool count. In addition, it was found that fees decreased with the change from POW to POS. Using data from January 2021 through December 2022,  \citet{KaraivanovZarifian2024} confirms that a higher block gas limit is associated with lower transaction fees. 
 Studying the period around the Dencun upgrade, \citet{Parketal.} finds that EIP-4844\footnote{\citet{eip4844}} increased rollup usage on Ethereum while lowering rollup transaction fees. 

\subsection{Layer 2 Networks}
\citet{GintingTPS2024} studies time to finality across the Mainnet, L2s (Arbitrum and Optimism), and Ethereum Virtual Machine (EVM) compatible blockchains (Avalanche, Binance Smart Chain, and Harmony). Ginting finds that Arbitrum has the fastest time to finality at 1.7 seconds for simple transactions and 2.1 seconds for smart contracts.

There is also a considerable amount of literature that uses simulations to analyze both fees and speed. Simulations in \citet{MELO2026108316} show that when 90\% of transactions are routed to L2s from the Mainnet, the throughput on the Mainnet can increase by up to 20\%.
\cite{DyadeLayer2Cost2025} show that EIP-4844 can reduce gas costs for L2 rollup submissions by up to 52\%.  Throughput can also nearly be doubled in larger blob sizes. They do find some saturation when blob sizes exceed 96kb. 

\subsection{Cross-chain Comparisons}
\citet{WIJAYA2025200} compare Solana and Ethereum using two submission patterns: sequential submission, where transactions are sent one after another, and parallel submission, where they are sent simultaneously in batches. They measure TPS, latency, and gas fees and find that Solana outperforms Ethereum across all three metrics in both modes.  \citet{Caparros} finds that lower costs on Arbitrum and Polygon attract smaller orders away from the Mainnet.

\citet{alizadeh2025solana} caution that speed can have drawbacks. They study the period from November 14, 2023, through March 18, 2024, and find that Solana has a transaction failure rate of approximately 20\%, whereas Ethereum has only 0.1\%. Solana also has a 7.6\% zero-value transfer rate, while Ethereum has a 0.66\% rate.

\subsection{Our Contribution}
Our paper contributes to this literature in three ways. First, we document the joint effects of the Dencun, Pectra, Fusaka, BPO, and gas limit upgrades on Mainnet transaction speed and fees from January 2024 through March 2026. To our knowledge, no prior work covers all these upgrades. We extend the cross-chain comparison literature by jointly studying the Mainnet, L2s, Solana, and Polygon. We also provide quantitative forecasts of Mainnet and L2 speeds under extrapolated trends and the EIP-7938 proposal in the February 2026 L1 Strawmap, which has not yet been empirically studied.

We then examine the effects of higher block gas limits and blob targets on fees.  Our analysis also predicts dates at which the Mainnet and Layer 2s will reach lower fees than Solana.

\section{Ethereum Network Improvements}
\label{sec:blobsandgaslimits}
The improvements covered in this section target two outcomes for the network: increasing TPS and decreasing fees. Section \ref{sec:Protocol Forks} covers the hard forks that introduced and expanded blobs; Section \ref{sec:blobinc} presents blob usage data. Section \ref{sec:gl} documents the increases in the gas limit. Section \ref{sec:ethtxfeesandcount} reports Mainnet transaction counts and fees. Section \ref{sec:upgrades} provides information on future upgrades.

\subsection{Protocol Forks}
\label{sec:Protocol Forks}

\noindent On March 13, 2024, at 01:55:35 UTC, epoch number: 269,568, Ethereum launched Cancun-Deneb, ``Dencun".\footnote{\url{https://ethereum.org/ethereum-forks/}} Cancun contains execution-layer improvements, while Deneb contains consensus-layer upgrades.  For our analysis, the key change was EIP-4844,\footnote{\citet{eip4844}} which introduced blobs. Blobs are cheaper than standard transactions because the data is only stored temporarily and not processed through Ethereum's Virtual Machine (EVM), the computation engine for the network.  This 
provides a cheaper way for L2s to post their transaction data to the Mainnet.\footnote{\url{https://chain.link/article/eip-4844-blob-storage}} The target number of blobs in Dencun was three, and the maximum number of blobs was six. These numbers have grown progressively since their introduction. 

Prague-Electra,``Pectra", was released May 7, 2025, at 10:05:11 UTC, epoch number 364,032.\footnote{\url{https://ethereum.org/ethereum-forks/}} The upgrade combines execution-layer changes (Prague) and consensus-layer changes (Electra).  The key updates for our analysis are EIP-7840\footnote{\citet{eip7840}} which simplified the process for blob updates, and EIP-7549\footnote{\citet{eip7549}} which increased the target number of blobs to six and the maximum to nine. The blob count increase under Pectra was explicitly intended to provide short-term TPS improvements.\footnote{\cite{eip7691}}

Following Pectra, Fulu-Osaka, ``Fusaka", was released December 3, 2025, at 21:49:11 UTC, epoch number 411,392. The name consists of Osaka, the execution-layer upgrade, and Fulu, the consensus-layer upgrade.\footnote{\url{https://ethereum.org/roadmap/fusaka/}} Under Fusaka, EIP-7935\footnote{\cite{EIP7935}} increased the gas limit from 36 million to 60 million, and  EIP-7594\footnote{\citet{eip7594}} introduced Peer Data Availability Sampling (PeerDAS). PeerDAS allows the nodes to confirm blob data is present by checking only a sample of it, rather than the full dataset. This reduces the workload on each node.

For L2 users, the intended outcome was a reduction in transaction fees and higher blob TPS.\footnote{\url{https://ethereum.org/roadmap/fusaka/peerdas/}} 
\subsection{Blob Capacity}
\label{sec:blobinc}

Fusaka also introduced a mechanism to rapidly expand blob capacity independently of other protocol changes.\footnote{\url{https://ethereum.org/roadmap/fusaka}}  These are called blob parameter-only (BPO) forks. These differ from traditional hard forks because BPOs adjust only a small set of blob-related parameters, such as the target and maximum numbers of blobs. Traditional hard forks require coordination and more extensive protocol changes.\footnote{\cite{eip7892}}

Since Fusaka went live, BPO-1 and BPO-2 have been activated. BPO-1
activated December 9, 2025, 14:21:11 UTC, at epoch 412,672, increasing the target blob count from six to 10, and the maximum from nine to 15. BPO-2 went live on January 7, 2026, 1:01:11 UTC, epoch
419,072. It increased the target blob count to 14 and the maximum to 21 blobs.\footnote{\url{https://www.optimism.io/blog/fusaka-is-live-scaling-optimism-and-the-superchain}}

We are interested in studying how the target blob count impacts both the TPS and transaction costs of the L2s. The target blob impacts the L2 fee calculation because when the blob usage exceeds the target number of blobs, the fee increases.\footnote{\cite{eip4844}} 

\begin{figure}[H]
	\centering
		\caption{Blob Usage and Targets}
		\label{fig:blob_target_max_avg}
        \begin{minipage}{0.97\linewidth}
        \begin{center}
			\includegraphics[width=0.97\textwidth]{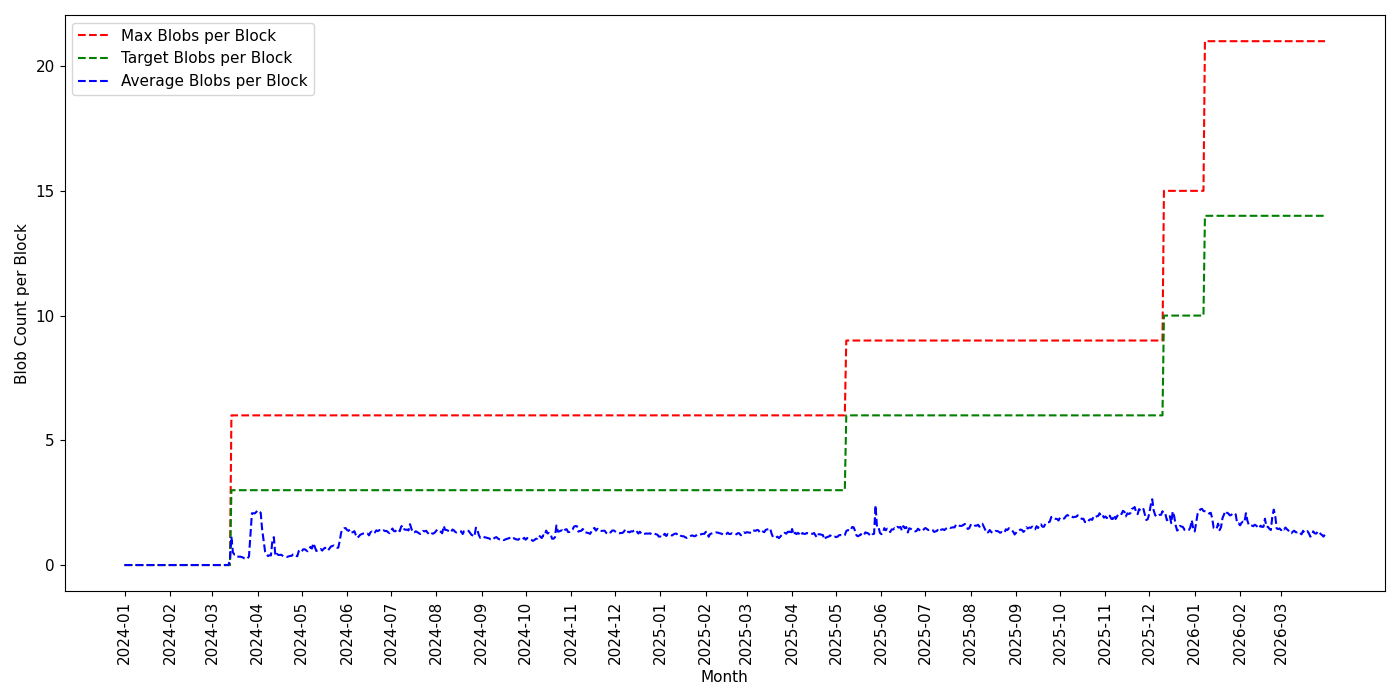} \\
		\end{center}
		\small
		\texttt{Note:} The blue line denotes the daily average blobs per block, green is the daily target blobs per block, and red is the daily maximum blobs per block. We use the complete transaction history of the Mainnet from Google BigQuery (\texttt{bigquery-publicdata.crypto\_ethereum}) and aggregate to the daily level. 
    \end{minipage}
\end{figure}

Figure \ref{fig:blob_target_max_avg} plots the daily number of target and maximum blobs per block and the average number of blobs in a block from January 2024 through March 2026.

 All three series begin at zero prior to the Dencun upgrade, at which point the target number of blobs jumps to three and the maximum to six. The target and maximum remain unchanged until the Pectra upgrade in May 2025, after which the target blob rises to six and the maximum to nine. The next jump is observed at BPO-1 in December 2025, raising the target to 10 and the maximum to 14, with BPO-2 shortly following in January 2026, raising the target further to 14 and the maximum to 21. 

Throughout the entire sample period, average blob usage, shown by the blue line, remains well below both the target and maximum, generally tracking around one to two blobs per block. While the average blob usage per block shows a slight uptrend, the difference between usage and the target and maximum blob counts continues to increase.

\begin{figure}[H]
	\centering
		\caption{Blob Usage Exceeding Targets}
		\label{fig:blob_target_demand}
        \begin{minipage}{0.97\linewidth}
        \begin{center}
			\includegraphics[width=0.97\textwidth]{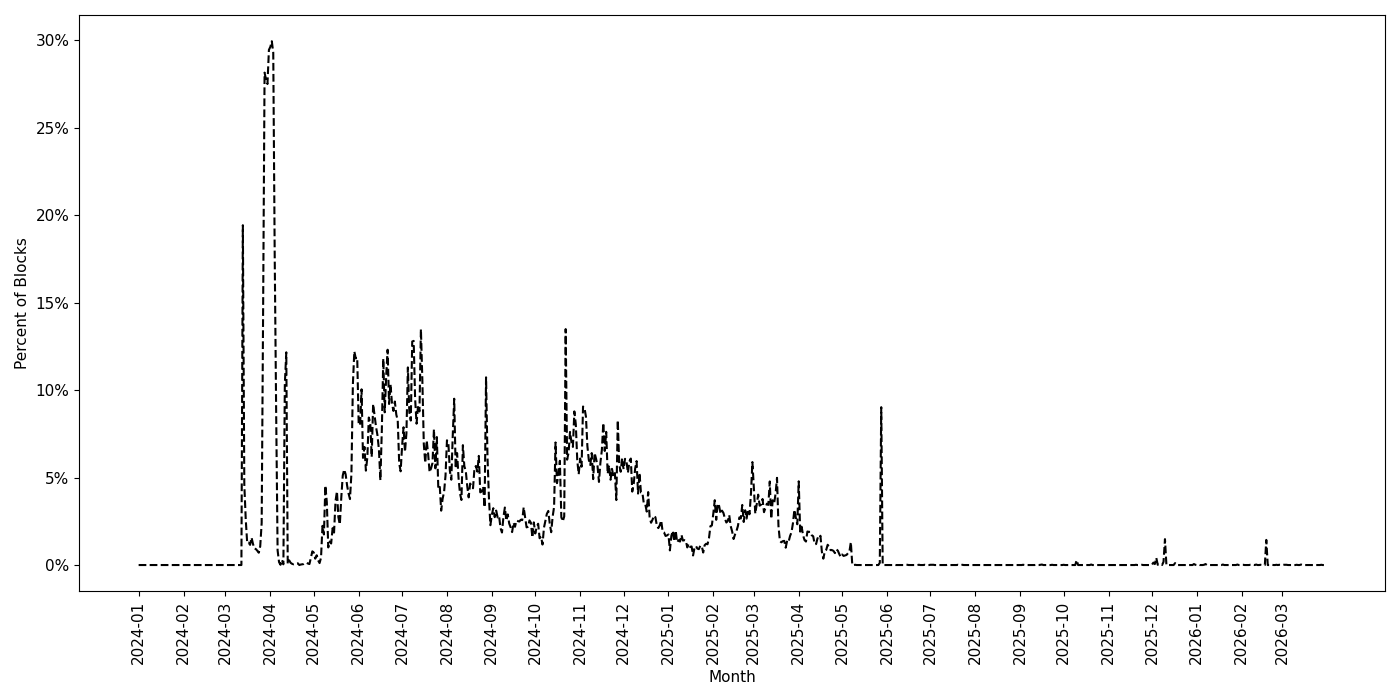} \\
		\end{center}
		\small
		\texttt{Note:} The black dotted line defines the daily percentage of blocks for which the number of blobs exceeds the target number of blobs per block. We use the complete transaction history of the Mainnet from Google BigQuery and aggregate to the daily level.
    \end{minipage}
\end{figure}

Figure \ref{fig:blob_target_demand} complements this result as it reports the percentage of blocks per day where the blob count exceeds the target. 

The proportion of blocks with the number of blobs exceeding the target declined over the sample. From  mid-2025 onward, the percentage of blocks exceeding the target fell to near zero and remained there through the end of the sample period.

\subsection{Gas Limit Data}
\label{sec:gl}
We are interested in studying how the gas limit affects both transaction costs and transaction speeds on the Mainnet. When a block's gas usage exceeds the target gas limit, defined as half of the gas limit, the following block incurs a fee increase.\footnote{\url{https://ethereum.org/developers/docs/gas/}} In terms of speed, increasing the gas limit increases the transaction count, enabling each block to handle more transactions. Ethereum Foundation developer, Marius van der Wijden, has documented this relationship.\footnote{\url{https://mariusvanderwijden.github.io/blog/2024/01/11/GasLimit/}} Consistent with this, \cite{nethermind2025gasbenchmark} finds that increasing gas limit increases TPS.

\begin{figure}[H]
	\centering
		\caption{Gas Limit Compared to Usage}
		\label{fig:eth_gas_boxplot}
        \begin{minipage}{0.97\linewidth}
        \begin{center}
			\includegraphics[width=0.97\textwidth]{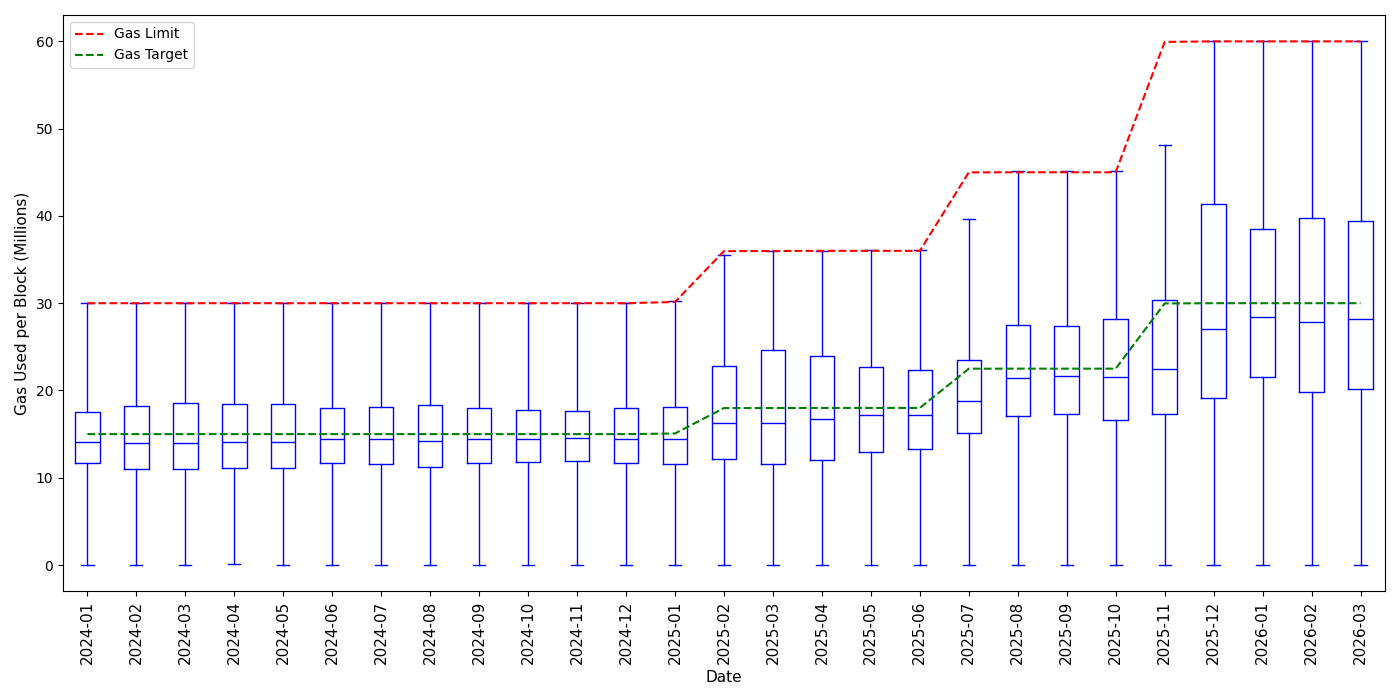} \\
		\end{center}
		\small
		\texttt{Note:} Each box represents the interquartile range (IQR) of monthly distributions of daily gas used per block, with the horizontal line inside each box indicating the median. The blue whiskers extend to the minimum and maximum observed values within each month. The red dashed line represents the monthly maximum of the daily median gas limit per block, and the green dashed line represents the gas target, defined as half of the gas limit. Daily data is obtained from our public Dune query: \url{https://dune.com/queries/7347447}.
    \end{minipage}
\end{figure}

Figure \ref{fig:eth_gas_boxplot} reports the high and low and interquartile range (IQR) of gas used per block within each month from January 2024 through March 2026, as well as the gas limit and the target gas limit. The IQR is defined as the difference between the 75th and 25th percentiles of daily gas used per block within each month. 

From January 2024 through January 2025, the gas limit remained flat at roughly 30 million, and the monthly distributions of gas used per block were narrow and stable, with the IQR averaging 6.6 million.

As the median gas limit increased to 36 million from February through June 2025, the average IQR widened to 10.9 million. The IQR remained at a similar level of 10.1 million during the 45 million gas limit period from July through October 2025. Following the increase to 60 million beginning in November 2025, the average IQR widened further to 18.3 million through March 2026.

While the medians in Figure \ref{fig:eth_gas_boxplot} are consistently below the target, under EIP-1559, it only requires a single block exceeding the target to increase the base fee for the following block. 

\begin{figure}[H]
	\centering
		\caption{Gas Usage Exceeding Targets}
		\label{fig:eth_gas_above_target}
        \begin{minipage}{0.97\linewidth}
        \begin{center}
			\includegraphics[width=0.97\textwidth]{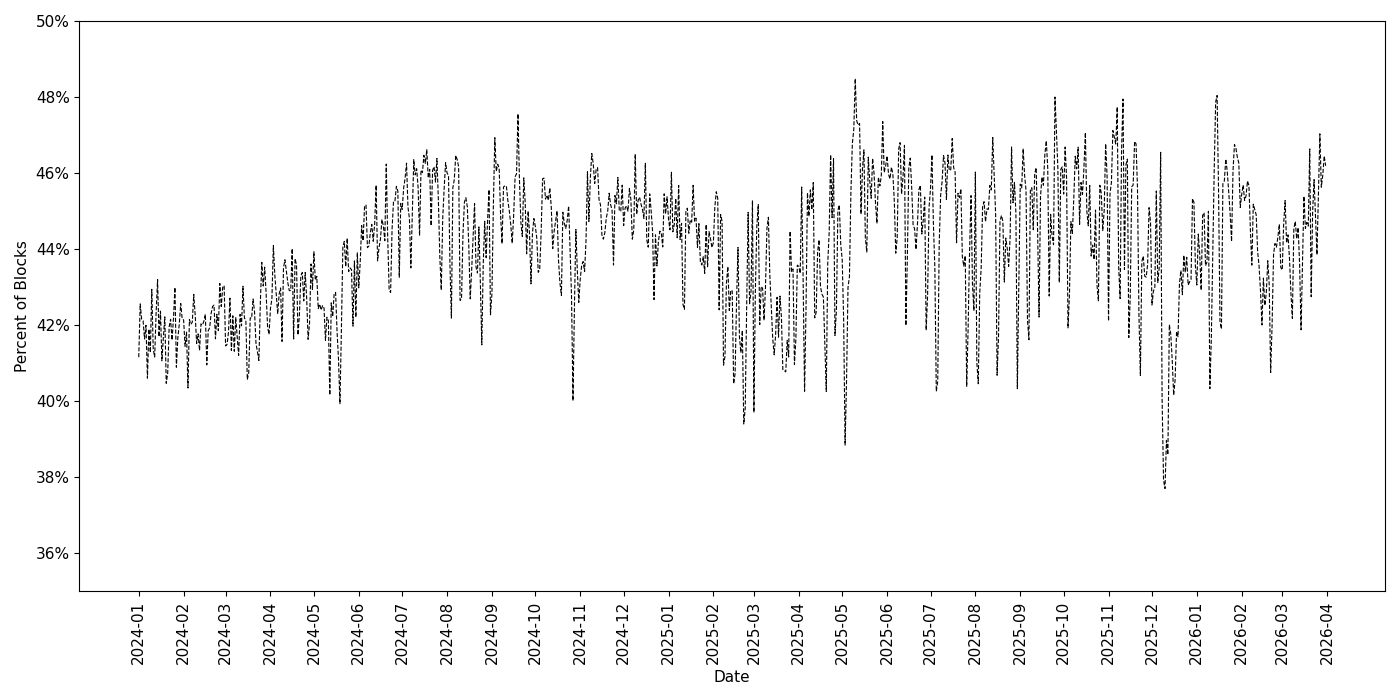} \\
		\end{center}
		\small
		\texttt{Note:} The black dotted line denotes the daily percentage of blocks in which gas used exceeded the target gas per block, defined as half the block gas limit. Data is obtained from our public Dune query: \url{https://dune.com/queries/7347447}.
    \end{minipage}
\end{figure}

Figure \ref{fig:eth_gas_above_target} reports the daily percentage of blocks in which gas used exceeded the target from January 2025 through March 2026.

Throughout our entire sample period, 37.7\% to 48.5\% of blocks exceeded the gas target on any given day. The percentage was lower and less volatile in early 2024, ranging from 39.9\% to 46.2\% from January 2024 through June 2024. There is no clear downward trend over the sample period, indicating that blocks exceeding the gas target persisted and remained consistent despite increases in the gas limit.

\subsection{Mainnet Transaction Information}
\label{sec:ethtxfeesandcount}
\begin{figure}[H]
	\centering
		\caption{Mainnet Monthly Transaction Counts}
		\label{fig:mainnet_monthly_transactions}
        \begin{minipage}{0.97\linewidth}
        \begin{center}
			\includegraphics[width=0.97\textwidth]{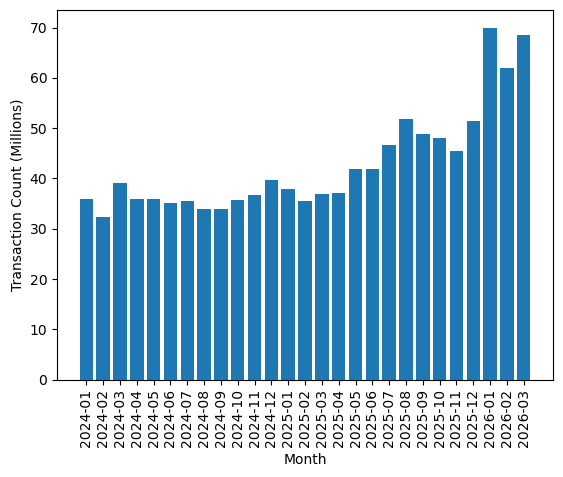} \\
		\end{center}
		\small
		\texttt{Note:} The blue bars denote the total monthly transaction counts on Ethereum. The data is obtained from Google BigQuery’s public Ethereum transaction data and aggregated monthly.
    \end{minipage}
\end{figure}

With the gas increases documented, we examine the transaction activity these changes have enabled on the Mainnet in Figure \ref{fig:mainnet_monthly_transactions}.

 Mainnet activity remained relatively flat between 32.3 and 39.6 million transactions per month from January 2024 through April 2025. Beginning in May 2025, transaction counts increased notably, rising to 70.01 million by January 2026 and remaining elevated through March 2026. 

\begin{figure}[H]
	\centering
		\caption{Mainnet Monthly Transaction Fees}
		\label{fig:ethereum_fees_split}
        \begin{minipage}{0.97\linewidth}
        \begin{center}
			\includegraphics[width=0.97\textwidth]{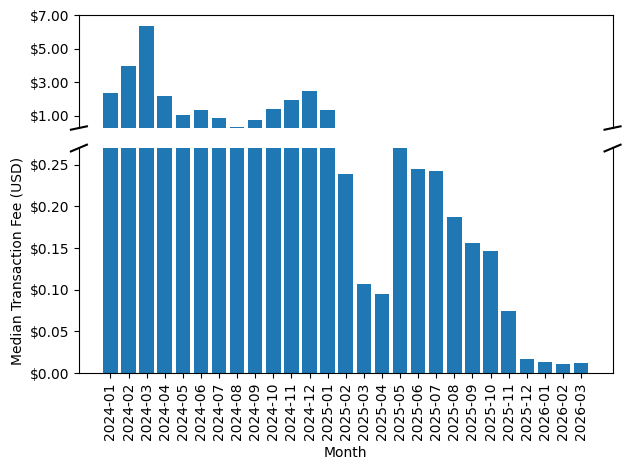} \\
		\end{center}
		\small
		\texttt{Note:} The blue bars denote the monthly median of daily median transaction fees in USD on Ethereum. Data for the Mainnet median transaction fee in ETH is obtained from Google BigQuery’s public Ethereum transaction data. The daily closing price of ETH is then collected from CoinGecko: \url{https://www.coingecko.com/en/coins/ethereum/historical_data} to convert fees to USD. The y-axis is broken to improve readability. The upper panel displays values from \$0.27 to \$7.00 and the lower panel displays values from \$0.00 to \$0.27.
    \end{minipage}
\end{figure}

Figure \ref{fig:ethereum_fees_split} presents Mainnet median monthly transaction fees.

Mainnet fees declined with each major gas limit increase. We will explore this relationship further in Section \ref{sec:EM}
\subsection{Roadmap for future upgrades}
\label{sec:upgrades}
In February 2026, the Ethereum Foundation introduced their L1 Strawmap, a roadmap of proposed protocol upgrades for the Mainnet.\footnote{\url{https://strawmap.org}}
The next protocool upgrade is expected to be Glamsterdam in 2026 Q3. This dynamics of this upgrade are expected to set the stage to reach 200 million gas limit floor once implemented.\footnote{\url{https://blog.ethereum.org/2026/05/02/soldogn-interop-recap}}

EIP-7938, currently proposed for inclusion in a future hard fork, would introduce a deterministic schedule that increases the gas limit by a factor of ten every two years for four years.\footnote{\cite{eip7938}} This trajectory aligns with the Strawmap's target of one gigagas per second.

\section{Polygon}
\label{sec:pol}
Polygon is an Ethereum Virtual Machine (EVM)-compatible side-chain that can run smart contracts and decentralized applications written for Ethereum.\footnote{\url{https://ethereum.org/developers/docs/scaling/sidechains/}} Operating as a sidechain, Polygon processes transactions on its own network and will periodically post a checkpoint to the Mainnet. This differs from L2s in that Polygon does not post its transaction data to the Mainnet to be validated.

\begin{figure}[H]
	\centering
		\caption{Polygon Monthly Transaction Counts}
		\label{fig:polygon_transaction_counts}
        \begin{minipage}{0.97\linewidth}
        \begin{center}
			\includegraphics[width=0.97\textwidth]{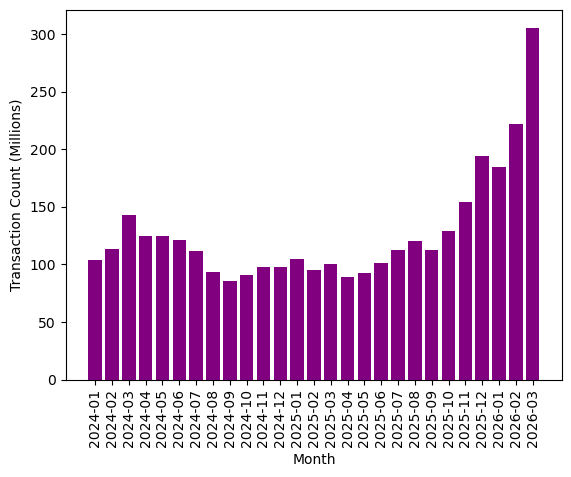} \\
		\end{center}
		\small
		\texttt{Note:} The purple bars denote the total monthly transaction counts on Polygon. The data is obtained from Google BigQuery's public Polygon transaction data (\texttt{bigquery-public-data.goog\_blockchain\_polygon\_mainnet\_us}) and aggregated monthly. 
    \end{minipage}
\end{figure}

Figure \ref{fig:polygon_transaction_counts} displays Polygon monthly transaction counts from January 2024 through March 2026. 

Transaction counts showed no sustained directional trend from January 2024 through April 2025, averaging approximately 106 million transactions per month. Beginning in May 2025, transaction counts trended upward, surpassing the prior sample high of 142.7 million by November 2025 and reaching 305.6 million by March 2026.

\begin{figure}[H]
	\centering
		\caption{Polygon Monthly Median Transaction Fees}
		\label{fig:polygon_median_fee_monthly_2024_2026}
        \begin{minipage}{0.97\linewidth}
        \begin{center}
			\includegraphics[width=0.97\textwidth]{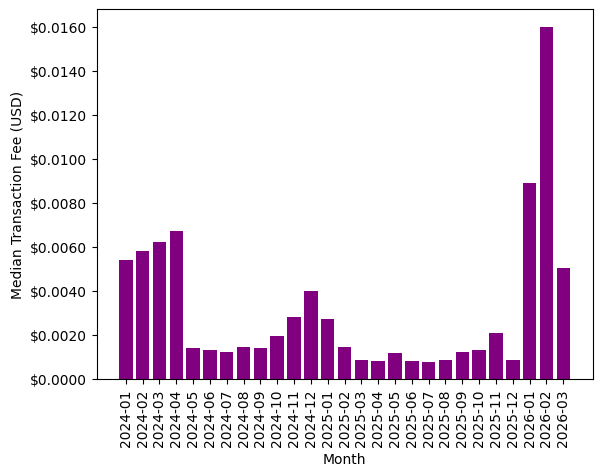} \\
		\end{center}
		\small
		\texttt{Note:} The purple bars denote the monthly median of daily median transaction fees in USD on Polygon. Transaction fees in POL, formerly MATIC, are obtained from Google BigQuery’s public Polygon transaction data. The daily closing
        price of POL is then collected from CoinGecko:     \url{https://www.coingecko.com/en/coins/polygon/historical_data}  to convert fees to USD.
    \end{minipage}
\end{figure}

Figure \ref{fig:polygon_median_fee_monthly_2024_2026} reports Polygon median monthly transaction fees in U.S. dollars.
Polygon had a sustained decline in fees beginning in May 2024.  Fees surged in the first quarter of 2026.  In February, total daily fees on Polygon surpassed Ethereum for the first time.  Industry analysis has focused on the prediction market Polymarket,\footnote{\url{https://polymarket.com/}} as the key driver.\footnote{\url{https://cryptorank.io/news/feed/39df8-polygon-fees-surpass-ethereum-polymarket}} 
\section{Solana}  
Solana serves as the primary performance benchmark for this paper. Solana is designed to handle high transaction speed on its base layer.

\begin{figure}[H]
	\centering
		\caption{Comparison of Solana Transactions}
		\label{fig:solana_vote_nonvote_transaction_counts}
        \begin{minipage}{0.97\linewidth}
        \begin{center}
			\includegraphics[width=0.97\textwidth]{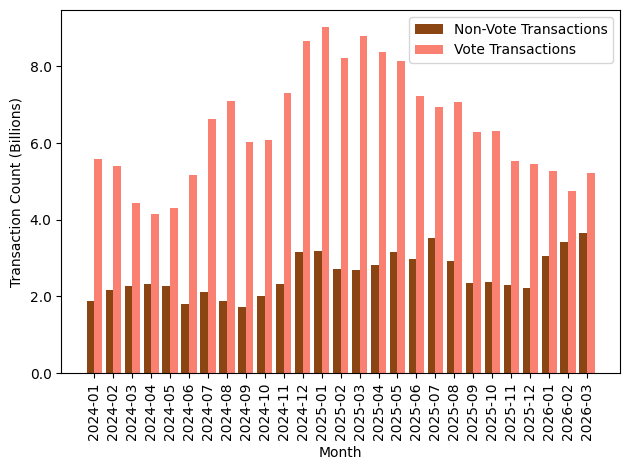} \\
		\end{center}
		\small
		\texttt{Note:} The brown bars denote the total non-vote monthly transaction counts on Solana. The pink bars denote the total vote monthly transaction counts on Solana. Daily transaction counts are obtained from our public Dune query: \url{https://dune.com/queries/6946173} and aggregated monthly.
    \end{minipage}
\end{figure}

Solana's actual speed is often overstated because a substantial portion of block space is required to validate transactions in Solana's consensus scheme.  Figure \ref{fig:solana_vote_nonvote_transaction_counts} displays Solana's monthly transaction counts, separated into vote and non-vote transactions, in billions from January 2024 through March 2026.

 Vote transactions make up an average of 71.5\% of total transactions in our sample. Solana's roadmap plans to open more block space for non-vote transactions through their Alpenglow upgrade. This upgrade, expected in 2026 Q3, will move all vote transactions off-chain.\footnote{\url{https://solana.com/news/solana-network-upgrades}}

\begin{figure}[H]
	\centering
		\caption{Solana Non-Vote Monthly Transaction Counts}
		\label{fig:solana_nonvote_trans_counts}
        \begin{minipage}{0.97\linewidth}
        \begin{center}
			\includegraphics[width=0.97\textwidth]{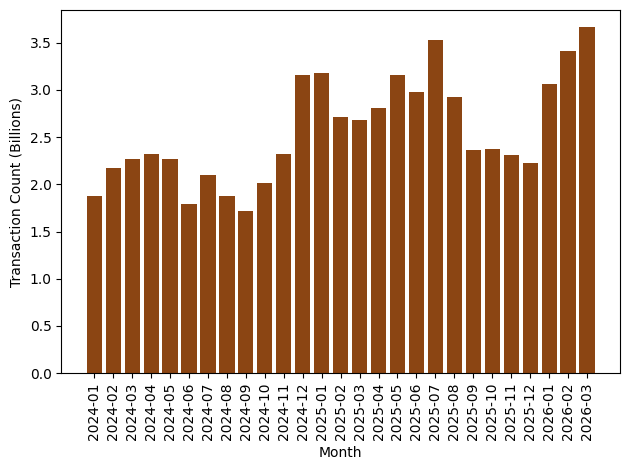} \\
		\end{center}
		\small
		\texttt{Note:} The brown bars denote the total non-vote monthly transaction counts on Solana. Daily transaction data are obtained from our public Dune query: \url{https://dune.com/queries/6946173} and aggregated monthly. 
    \end{minipage}
\end{figure}

 We will focus our comparisons across the blockchains on the non-vote transactions which we plot in Figure \ref{fig:solana_nonvote_trans_counts}.

Transaction counts were relatively stable throughout the sample period, ranging between 1.72 billion transactions per month in September 2024 to 3.66 billion in March 2026. A notable decline is observed beginning in August 2025, falling from approximately 2.92 billion to approximately 2.23 billion by December 2025, before recovering to approximately 3.66 billion by March 2026

\begin{figure}[H]
	\centering
		\caption{Solana Monthly Median Transaction Fees}
		\label{fig:solana_median_fee_monthly_2024_2026}
        \begin{minipage}{0.97\linewidth}
        \begin{center}
			\includegraphics[width=0.97\textwidth]{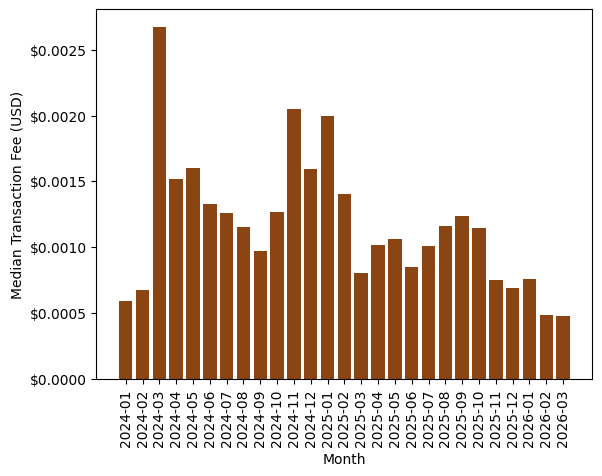} \\
		\end{center}
		\small
		\texttt{Note:} The brown bars denote the monthly median of daily median transaction fees in USD on Solana. Transaction fees are obtained from our Dune query: \url{https://dune.com/queries/6946173}.
    \end{minipage}
\end{figure}

Figure \ref{fig:solana_median_fee_monthly_2024_2026} presents Solana median monthly transaction fees.  Solana again is a benchmark with the lowest fees of any chain other than Optimism.

Fees on Solana show a sustained downtrend from March 2024, but fees fell below levels from the first two months of 2024 only in February and March of 2026.

\section{L2 Networks}
\label{sec:l2n}
There are two different classes of L2 networks: optimistic rollups and Zero-knowledge (ZK) rollups. All L2s execute transactions off-chain and post the compressed transaction data to the Mainnet. The difference between the two classes lies in their verification approach.

For optimistic rollups, it is assumed that the transactions posted by the L2 are valid. There is a 7-day window during which a transaction can be challenged for review by the Mainnet. Due to the immediate validity assumption, optimistic rollups can achieve higher TPS than ZK-rollups.

Rather than assume validity upon posting data to the Mainnet, ZK-rollups provide cryptographic validity proofs for each transaction batch posted to the Mainnet, which must be verified. This method is much more computationally intensive, limiting the TPS count.

\subsection{Transaction Comparison}

\begin{figure}[H]
	\centering
		\caption{L2 TPS Comparison}
		\label{fig:optimistic_zkrollup_tps}
        \begin{minipage}{0.97\linewidth}
        \begin{center}
			\includegraphics[width=0.97\textwidth]{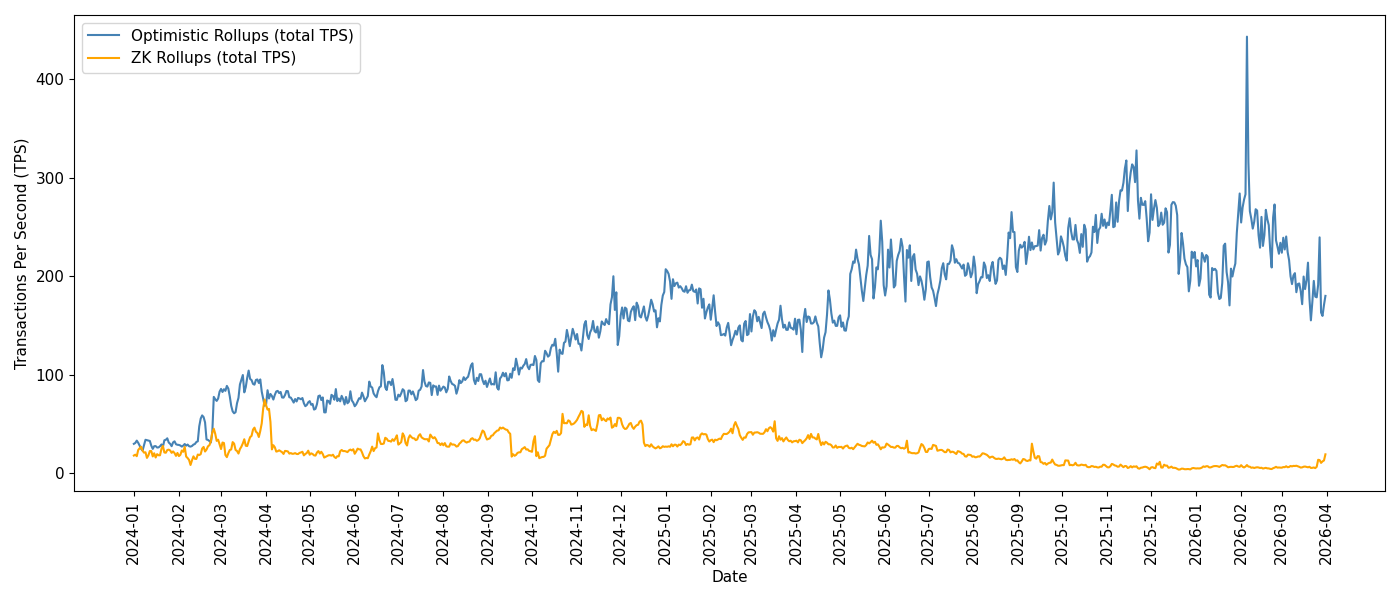} \\
		\end{center}
		\small
		\texttt{Note:} The orange line denotes the average daily TPS of ZK-Rollups. The blue line denotes the average daily TPS of Optimistic Rollups. TPS is calculated using the sum of daily transaction counts divided by the number of seconds in a day. The following are used in calculating Optimistic Rollup TPS: Arbitrum, Base, Blast, Bob, Boba, Hemi, Ink, Nova, Optimism, Shape, Superseed, Unichain, Worldchain, and B3. The following are used in calculating ZK-Rollup TPS: Abstract, Linea, Scroll, Taiko, ZkEVM, and ZkSync. The sample is restricted to L2s that use ETH as their gas fee token. Data is obtained from our public Dune query: \url{https://dune.com/queries/6962020}.
    \end{minipage}
\end{figure}

Figure \ref{fig:optimistic_zkrollup_tps} plots daily TPS for optimistic and ZK-rollups from January 2024 through March 2026. Optimistic rollups grew steadily over the period, rising from 29.67 TPS in January 2024 to 195.22 TPS for March 2026. ZK-rollups followed a different trajectory, reaching 33.57 TPS in March 2024 before declining sharply to around 7.82 TPS in March 2026.

\begin{figure}[H]
	\centering
		\caption{L2 Transaction Counts}
		\label{fig:layer2_monthly_transactions}
        \begin{minipage}{0.97\linewidth}
        \begin{center}
			\includegraphics[width=0.97\textwidth]{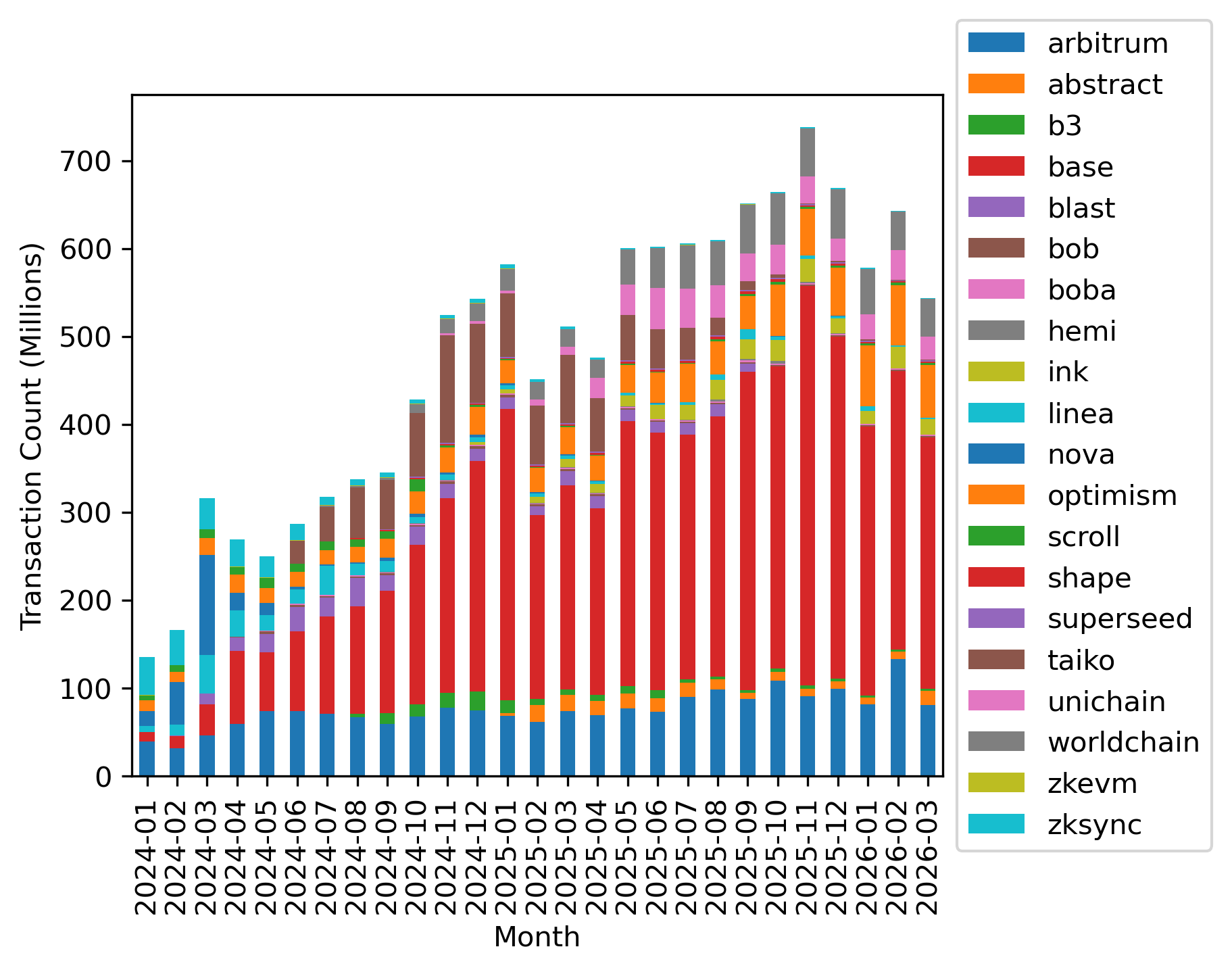} \\
		\end{center}
		\small
		\texttt{Note:} Each colored segment denotes the total monthly transaction counts for an individual L2 network, stacked to show the total monthly transaction counts across all L2s. The L2 networks included are: Arbitrum, Abstract, B3, Base, Blast, Bob, Boba, Hemi, Ink, Linea, Nova, Optimism, Scroll, Shape, Superseed, Taiko, Unichain, Worldchain, ZkEVM, and ZkSync. Daily data is obtained from our public Dune query: \url{https://dune.com/queries/6962020} and aggregated monthly.
    \end{minipage}
\end{figure}

 Figure \ref{fig:layer2_monthly_transactions} includes all 20 L2 networks that collect fees in ETH.  We do not include L2s such as Celo\footnote{\url{https://l2beat.com/scaling/projects/celo}} that do not price fees in ETH.  Of the 20 L2 networks in our sample, nine were active from January 1, 2024. 11 entered throughout the sample period, with the last network, Ink, joining in December 2024. 

The combined L2 networks have outperformed the Mainnet in transaction count through the entire period. Moreover, Arbitrum and Base individually exceeded the Mainnet's monthly transaction count for most of the sample period, with Base surpassing the Mainnet from April 2024 onwards, and Arbitrum from January 2024 onwards.

Total L2 transactions grew from 135.6 million in January 2024 to 543.8 million in March 2026. Base grew quickly to become the dominant network throughout the period and is the primary driver of overall growth. Base also leads in DEX trading, as we show in Figure \ref{fig:dex_volume_share_2026Q1}, where Base has a larger volume than all the other L2s. Since April 2025, Arbitrum has had the second-highest transaction count among the L2s.  In Q1 2026, Arbitrum, Base, and Optimism are the three networks with the highest transaction counts.

Although Arbitrum, Base, and Optimism are optimistic rollups and EVM compatible, their schemes are different. Optimism was publicly launched on December 16, 2021.\footnote{\url{https://messari.io/project/optimism/profile}} In 2024, Optimism introduced OP Stack, a shared software system that was adopted by Base, Unichain and Zora, among others. Blockchains built on OP Stack form a connected network known as the Superchain.\footnote{\url{https://eco.com/support/en/articles/11779236-what-is-the-op-stack-architecture-and-superchain-explained}} 

Base was developed by Coinbase and launched in August 2023. Base has access to Coinbase's existing users, enabling a smoother transition for users moving from Coinbase to Ethereum L2 applications. Specifically, decentralized applications (dApps) built on Base have access to Coinbase's existing products and developer tools, including wallet infrastructure. Base was the first major network built using OP Stack, but as of May 2026 it has adopted its own software stack.\footnote{\url{https://blog.base.dev/next-chapter-for-base-chain-1}} In addition to this change, Base has moved away from purely optimistic proofs to using ZK and Trusted Execution Environment proofs.\footnote{\url{https://blog.base.dev/multiproofs-on-base}}

Arbitrum became publicly accessible on August 31, 2021.\footnote{\url{https://messari.io/project/arbitrum/profile}} It consists of two networks, Arbitrum One and Arbitrum Nova.  We only include Arbitrum One here because it handles more than 95\% of the volume.\footnote{\url{https://defillama.com/chain/arbitrum-nova}}  On January 8, 2026, Arbitrum introduced a fee-smoothing mechanism intended to reduce price volatility during periods of high demand.
 
\section{Blockchain Comparison}
\label{sec:blockchain_compare}
We now quantify the differences between the Mainnet, its L2s, Solana, and Polygon by comparing DEX volume and TPS from January 2024 through March 2026.

\subsection{Volume}
\label{sec:dex}
\begin{figure}[H]
	\centering
		\caption{DEX USD Transfer Value Comparison 2026 Q1}
		\label{fig:dex_volume_share_2026Q1}
        \begin{minipage}{0.97\linewidth}
        \begin{center}
			\includegraphics[width=0.97\textwidth]{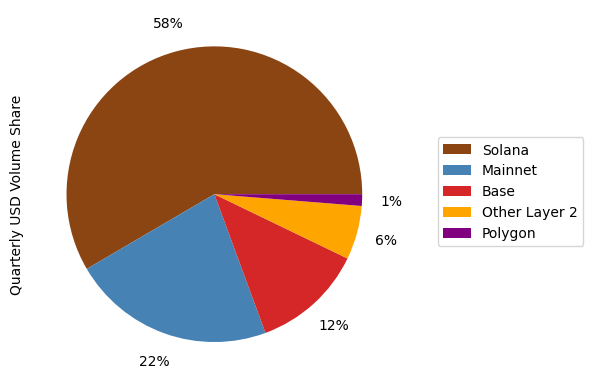} \\
		\end{center}
		\small
		\texttt{Note:} Each slice of the pie chart denotes the share of total USD-valued DEX transfer volume for a given blockchain. Brown denotes Solana, blue denotes the Mainnet, red denotes Base, orange denotes other L2s, and purple denotes Polygon. The ``other L2" category includes any L2 with less than \$50 billion in DEX volume. Data is obtained from our public Dune query: \url{https://dune.com/queries/6946454}. 
    \end{minipage}
\end{figure}

To analyze trading activity, we aggregate 2026 Q1 DEX volumes across all trading pairs and convert them into USD equivalents.  This measure captures the total dollar value of DEX  trades executed on each blockchain. We compute this metric for the Mainnet, Base, Solana, and Polygon. We group all the other L2s, which each have less than \$50 billion of DEX volume, into another group, `Other Layer 2'. Figure \ref{fig:dex_volume_share_2026Q1} therefore represents the distribution of trading volume measured in USD across DEX on each blockchain.

Solana leads with 58\% of total volume, followed by Ethereum at 22\%. Base has the greatest volume percentage compared to all other L2s at 12\%, leaving other L2s to make up 6\% of volume. Polygon has the lowest volume at 1\%. Combined, Ethereum and its L2 networks account for 40\% of DEX volume. 

\subsection{Transactions Per Second}
\label{sec:solspeed} 
We begin by comparing 2026 Q1 TPS to the same quarter back in 2024. To study L2 speed, we aggregate the daily TPS across all L2 networks introduced in Section~\ref{sec:l2n} to obtain a combined L2 TPS measure. We refer to this as L2 Total TPS.

Table \ref{tab:speed_summary} reports the 2024 Q1, 2024, 2025, and 2026 Q1 TPS values for Ethereum, L2 Total, Polygon, and Solana.

\begin{table}[H]
\centering
\caption{Network Speed: Average Daily Transactions per Second}
\label{tab:speed_summary}
\begin{minipage}{0.97\linewidth}
\begin{center}
\begin{tabular}{lrrrr}
\hline
\textbf{Network} & \textbf{2024 Q1} & \textbf{2024} & \textbf{2025} & \textbf{2026 Q1} \\
\hline
Mainnet    & 13.67  & 13.59  & 16.61    & 25.78    \\
L2 Total   & 78.60  & 124.00 & 227.08   & 226.92   \\
Solana     & 804.72 & 819.28 & 1,054.77 & 1,303.46 \\
Polygon    & 45.72  & 41.32  & 44.57    & 91.51    \\
\hline
\end{tabular}
\end{center}
\small
\texttt{Note:} Annual values are computed as the average of daily TPS across each calendar year. 2024 Q1 and 2026 Q1 values cover January 1 through March 31 of their respective years. 
\end{minipage}
\end{table}
 Solana remains the fastest network throughout the sample, averaging 804.72 TPS in 2024 Q1 and rising to 1,303.46 TPS by 2026 Q1. The L2 ecosystem has closed the gap with Solana, rising from 78.60 TPS in 2024 Q1 to 226.92 TPS in 2026 Q1.  Polygon doubled its TPS rising from 45.72 in 2024 Q1 to 91.51 TPS in 2026 Q1. The Mainnet grew almost as quickly from 13.67 TPS in 2024 Q1 to 25.78 TPS in 2026 Q1. 

\begin{figure}[H]
	\centering
		\caption{Blockchain Transaction Speeds}
		\label{fig:blockchain_tps_comparison}
        \begin{minipage}{0.97\linewidth}
        \begin{center}
			\includegraphics[width=0.97\textwidth]{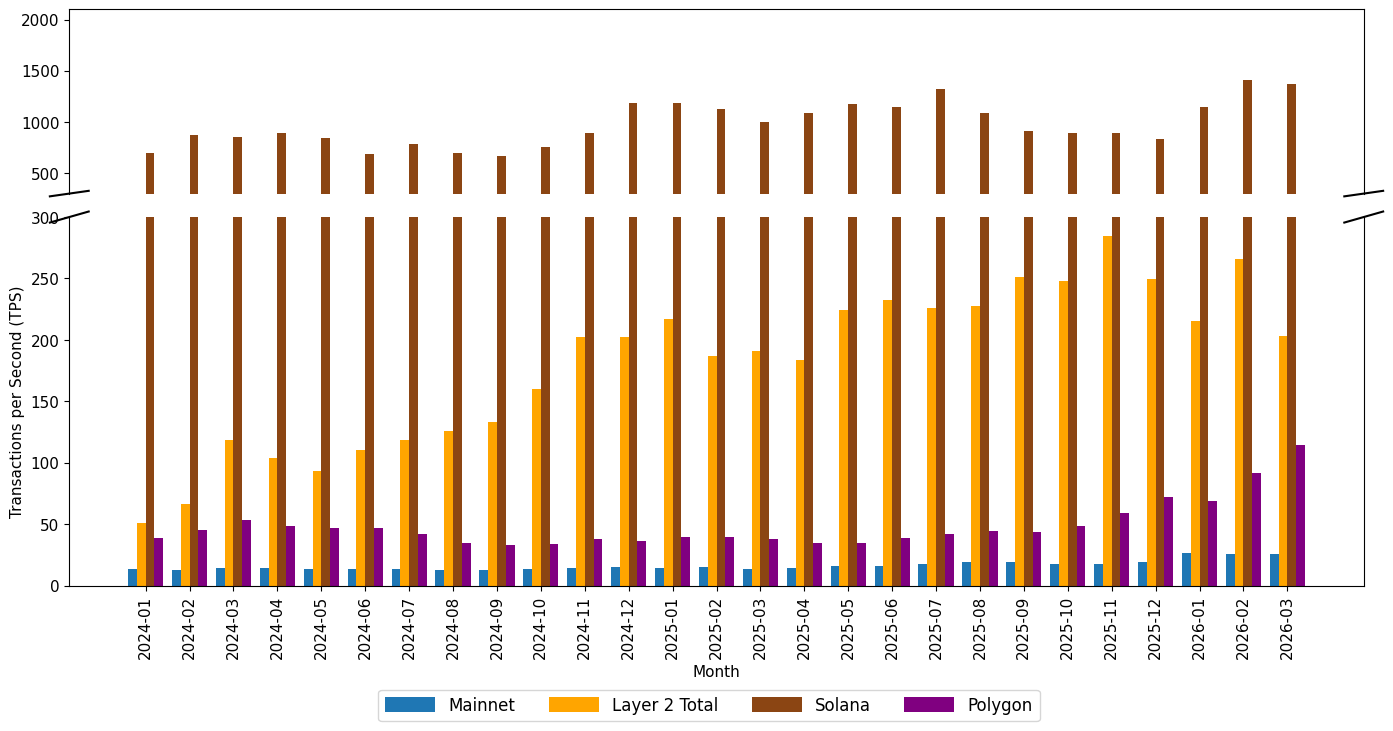} \\
		\end{center}
		\small
		\texttt{Note:} Each set of bars denotes the monthly TPS for a given network. TPS is calculated using the sum of monthly transaction counts divided by the number of seconds in a a month.  The blue bars denote the Mainnet, the orange bars denote the Layer 2 Total, the brown bars denote Solana, and the purple bars denote Polygon.  Mainnet and Polygon data are obtained from their respective public transaction data on Google BigQuery. Solana data is obtained from our public Dune query: \url{https://dune.com/queries/6946173}. L2 data is obtained from our public Dune query: \url{https://dune.com/queries/6962020}. All data was collected at the daily level and aggregated monthly.
    \end{minipage}
\end{figure}

Figure \ref{fig:blockchain_tps_comparison} visualizes the TPS of all networks monthly from January 2024 through March 2026. 

All three networks show uptrends in TPS.  We will extrapolate time trends in Solana.  For the Mainnet, we will link TPS improvements to increases in the gas limit.  For the Layer 2s, we will model their trend growth using blobs.

\subsection{Transaction Fees}

To present a representative fee measure, we focus on the L2 Leaders, defined as the mean of daily median fees across Arbitrum, Base, and Optimism. These three networks account for 79.39\% of total L2 transactions and 18\% of total DEX volume in Q1 2026. 

Table \ref{tab:median_fees} reports the median of daily median transaction fees in USD for Mainnet, Solana, L2 Leaders, and Polygon across the sample period.
\begin{table}[H]
\centering
\caption{Median Transaction Fees in USD}
\label{tab:median_fees}
\begin{threeparttable}
\begin{tabular}{lcccc}
\toprule
\textbf{Network} & \textbf{2024 Q1} & \textbf{2024} & \textbf{2025} & \textbf{2026 Q1} \\
\midrule
Mainnet & \$3.786300 & \$1.736066 & \$0.164527 & \$0.012283 \\
Solana & \$0.000687 & \$0.001288 & \$0.001037 & \$0.000496 \\
\\
Arbitrum & \$0.134343 & \$0.006011 & \$0.002957 & \$0.002183 \\
Base & \$0.149274 & \$0.007602 & \$0.001658 & \$0.001634 \\
Optimism & \$0.136594 & \$0.007140 & \$0.000261 & \$0.000027 \\
\midrule
L2 Leaders & \$0.180219 & \$0.051259 & \$0.002305 & \$0.001512 \\
\\
Polygon & \$0.005742 & \$0.002503 & \$0.001070 & \$0.006868 \\
\bottomrule
\end{tabular}
\begin{tablenotes}[flushleft]
\item
\texttt{Note:} Fees are the medians of daily medians across each calendar year. 2024 Q1 and 2026 Q1 values cover January 1 through March 31 of their respective years. L2 Leaders consists of the average of the daily medians of Arbitrum, Base, and Optimism.
\end{tablenotes}
\end{threeparttable}
\end{table}

The L2 Leaders declined continuously across the sample, falling from \$0.180219 in 2024 Q1 to \$0.001512 in 2026 Q1. Mainnet fees observed a sharp decline across the sample, falling from an average of \$3.786300 in 2024 Q1 to \$0.012283 in 2026 Q1. Polygon was the only network to see an increase in fees, which occurred in 2026 Q1. Solana fees remained consistently low throughout the sample, with a median of \$0.000687 in 2024 Q1 and declining further to \$0.000496 in 2026 Q1.
\begin{comment}
The gap between the Mainnet and Solana was larger in transaction speed than in transaction fees in 2026 Q1, with a 50-fold TPS gap compared to a 27-fold fee gap.
\end{comment}
Although the gap between the Mainnet and Solana fees has narrowed substantially, Solana continues to offer the lowest transaction costs.

\begin{figure}[H]
	\centering
		\caption{L2 Transaction Fees}
		\label{fig:layer2_fees_comparison}
        \begin{minipage}{0.97\linewidth}
        \begin{center}
			\includegraphics[width=0.97\textwidth]{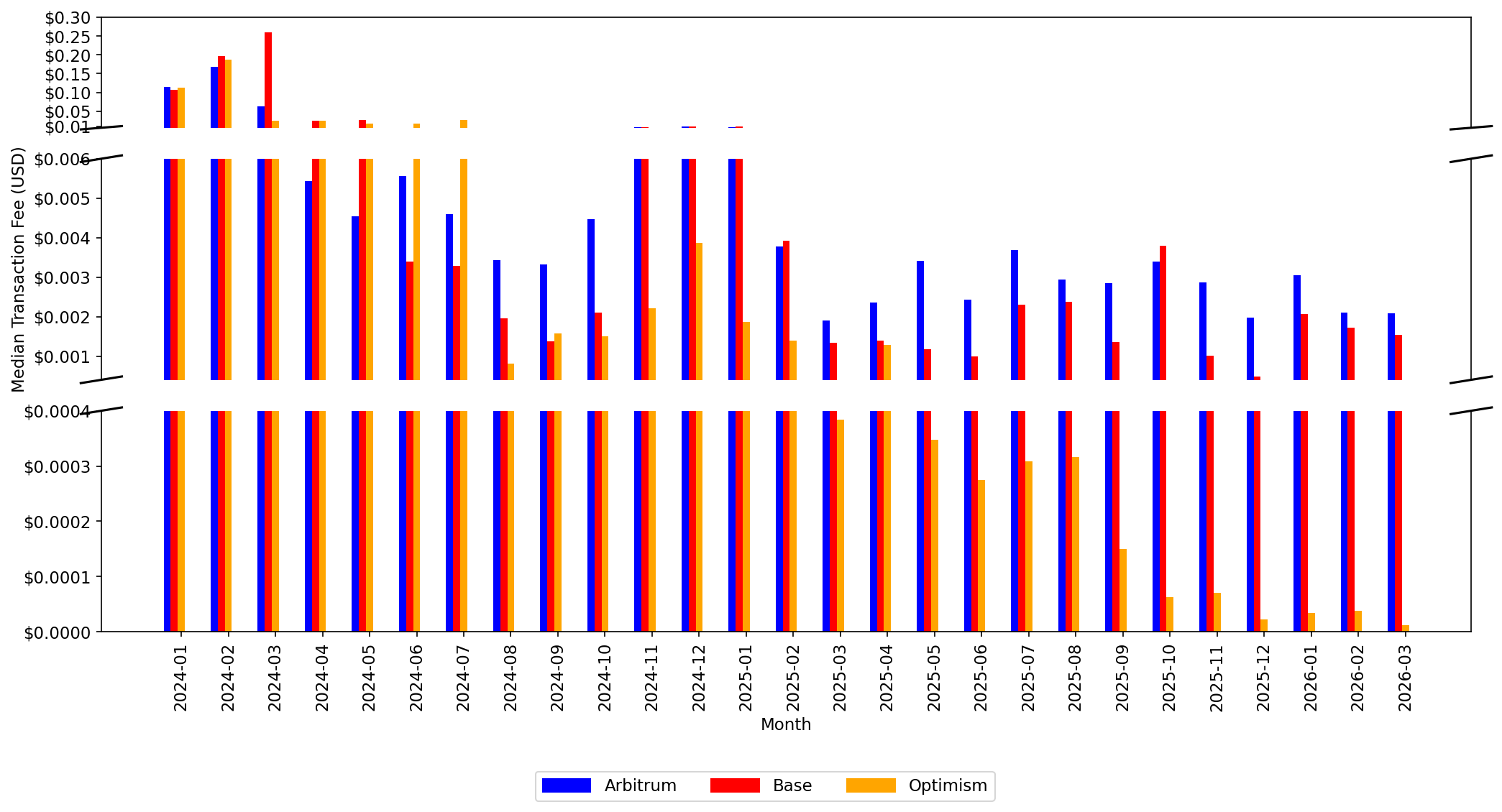} \\
		\end{center}
		\small
		\texttt{Note:} Each set of bars denotes the monthly median of daily median transaction fees in USD for a given network. Blue denotes Arbitrum, red denotes Base, and orange denotes Optimism. The y-axis is broken into three panels to improve readability. The upper panel displays values from \$0.006 to \$0.30, the middle panel displays values from \$0.0004 to \$0.006, and the lower panel displays values from \$0.0000 to \$0.0004. Daily data for the Arbitrum, Base, and Optimism are obtained from our public Dune query: \url{https://dune.com/queries/6962020}.
    \end{minipage}
\end{figure}

Figure \ref{fig:layer2_fees_comparison} displays median transaction fees in USD for Arbitrum, Base, and Optimism from January 2024 through March 2026. 

Following the Dencun upgrade in March 2024, all three networks experienced fee declines. Fees spiked again in late 2024 before falling further after the Pectra upgrade in May 2025, after which all three networks converged to their lowest levels of the sample period.

Optimism became consistently below Arbitrum in August 2024 and Base in October 2024. It remains, on average, 78.1\% below Abitrum and 75.2\% below Base.

Since May 2025, Optimism's median transaction fees have fallen below Solana's, averaging 81.7\% lower across the eleven months. All three L2s have consistently been below Mainnet fees for the entire period.

\begin{figure}[H]
	\centering
		\caption{Ethereum Competitor Transaction Fees}
		\label{fig:blockchain_fee_comparison}
        \begin{minipage}{0.97\linewidth}
        \begin{center}
			\includegraphics[width=0.97\textwidth]{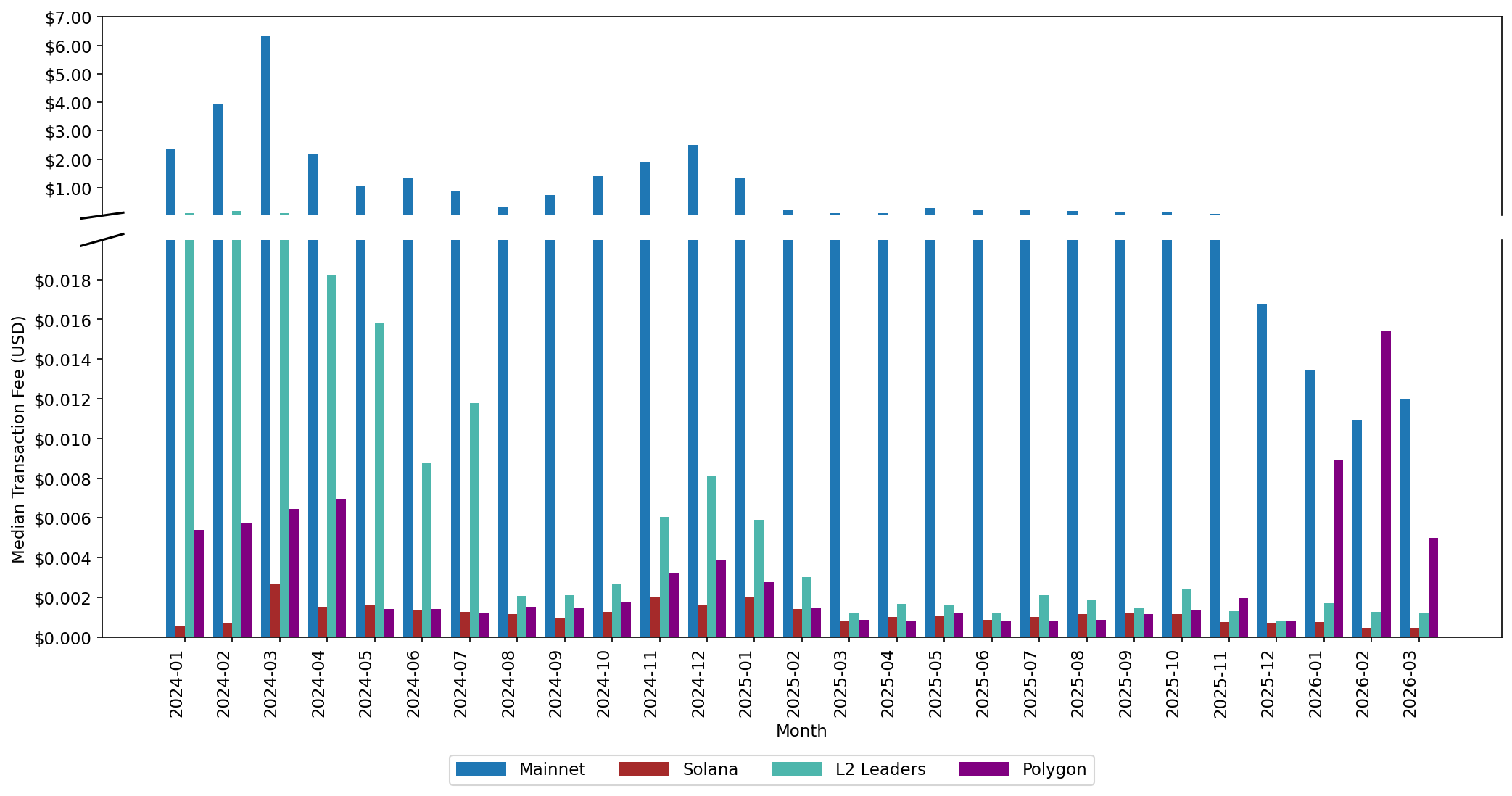} \\
		\end{center}
		\small
		\texttt{Note:} Each set of bars denotes the monthly median of daily median transaction fees in USD for a given network. Blue denotes the Mainnet, red denotes Solana, teal denotes L2 Leaders, and purple denotes Polygon. The y-axis is broken to improve readability. The upper panel displays values from \$0.02 to \$7.00 and the lower panel displays values from \$0.00 to \$0.02. Solana data is obtained from our public Dune query: \url{https://dune.com/queries/6946173}. L2 data is obtained from our public Dune query: \url{https://dune.com/queries/6962020}. Data for the Mainnet daily median transaction fee in ETH is obtained from Google BigQuery’s public Ethereum transaction data which we convert to USD using CoinGecko closing prices.
    \end{minipage}
\end{figure}

Figure \ref{fig:blockchain_fee_comparison} shows the monthly median transaction fees in USD for the Mainnet, the L2 Leaders, Solana, and Polygon from January 2024 through March 2026.

Each increase in the target blob count was associated with lower L2 transaction fees. Before the introduction of blobs through the Dencun upgrade, the L2 Leader fee was \$0.2107. The introduction of three target blobs under Dencun reduced this by 95.5\% to \$0.0095. Subsequent increases to six blobs under Pectra and 10 blobs under BPO-1 further reduced the fees to \$0.0019 and \$0.0013. The most recent increase to 14 blobs under BPO-2 saw fees rise slightly to \$0.0015.

As reported in Table \ref{tab:median_fees}, Solana provided,on average, the lowest median transaction fees across our sample period. However, there were seven months when Polygon had lower median fees than Solana, during the second and third quarters of both 2024 and 2025. During these months, Polygon was on average 10.9\% below Solana.

\section{Transactions Per Second Forecasts}
\label{sec:tps_forecast}
The previous sections study the historical differences in transaction speed and cost across the Mainnet and its competitors. We now examine how the gas limit affects Mainnet TPS and how the target blob count affects L2 TPS. We extend this study to include forecasts for both Mainnet and L2 TPS and compare the forecasts to our projected Solana speeds.

\subsection{Transactions Per Second on the Ethereum Mainnet}
\label{sec:Transaction Per Second on the Mainnet}
To study the impact of the gas limit on Mainnet TPS, we begin our sample in January 2025 when the gas limit began to increase toward 35 million.

Table \ref{tab:gas_trend} reports the estimates from a linear time trend regression of the gas limit over the period January 1, 2025 through March 31, 2026. The time trend coefficient of $7.44\times10^{4}$ indicates that the gas limit increases by 2.2 million units per month under the current growth trajectory.

\begin{table}[H]
\centering
\caption{Gas Limit Time Trend}
\label{tab:gas_trend}
\begin{threeparttable}
\begin{tabular}{lccccc}
\toprule
 & \multicolumn{5}{c}{Dependent variable: Gas Limit} \\
\cmidrule(lr){2-6}
 & Coef. & Std. Err. & $z$ & $p$-value & [0.025, 0.975] \\
\midrule
Constant 
& $2.78\times10^{7}$ & $6.36\times10^{5}$ & 43.78 & 0.000 & [$2.66\times10^{7}$,\;$2.91\times10^{7}$] \\
Time Trend ($t$)
& $7.44\times10^{4}$ & 2660.67 & 27.98 & 0.000 & [$6.92\times10^{4}$,\;$7.97\times10^{4}$] \\
\midrule
Observation Period & \multicolumn{5}{c}{2025-01-01 through 2026-03-31} \\
Adjusted $R^2$ & \multicolumn{5}{c}{0.88} \\
\bottomrule
\end{tabular}
\begin{tablenotes}[flushleft]
\item
\texttt{Note:} Gas limit is defined as the median daily gas limit per block. Standard errors are heteroskedasticity and autocorrelation consistent (HAC) with lag length selected automatically by a Bartlett kernel.
\end{tablenotes}
\end{threeparttable}
\end{table}

\begin{figure}[H]
	\centering
		\caption{Gas Limit: Observed vs Predicted}
		\label{fig:mainnet_gaslimit_timetrend}
        \begin{minipage}{0.97\linewidth}
        \begin{center}
			\includegraphics[width=0.97\textwidth]{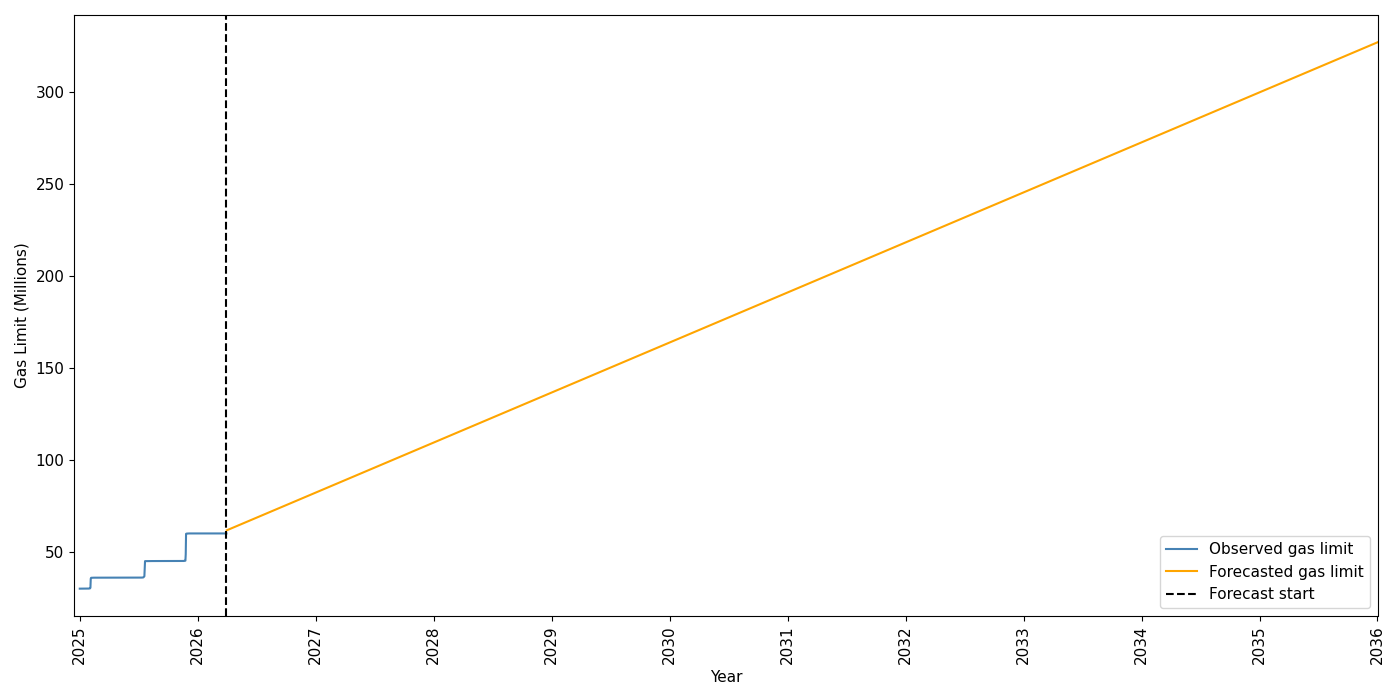} \\
		\end{center}
		\small
		\texttt{Note:} The blue line denotes the daily median gas limit from 2025-01-01 through 2026-03-31. The black vertical line indicates the beginning of our forecast. The orange line represents our extrapolated gas limit using Equation \ref{eq:gaslim}. Historical data for the gas limit is obtained from our public Dune query: \url{https://dune.com/queries/7347447}. 
    \end{minipage}
\end{figure}

Applying the results from Table \ref{tab:gas_trend}, we first utilize  this forecasting model for the gas limit,
\begin{equation}
\text{Gas Limit}_{t+\tau} = 2.78\times10^{7} + 7.44\times10^{4}\cdot(t+\tau)
\label{eq:gaslim}
\end{equation}
where $t = 1, 2, \ldots, T$ with $t=1$ corresponding to 2025-01-01 and $T=454$ corresponding to 2026-03-31, and $\tau \geq 0$ indexing days into the forecast horizon. Figure \ref{fig:mainnet_gaslimit_timetrend} presents the forecast of gas limit values until 2036.

Having established the trajectory of the gas limit under the current growth trend, we now determine the impact that the gas limit has on the Mainnet TPS. Table \ref{tab:tps_gas} 
reports the regression results.

\begin{table}[H]
\centering
\caption{Effect of Gas Limit on Ethereum Mainnet}
\label{tab:tps_gas}
\begin{threeparttable}
\begin{tabular}{lccccc}
\toprule
 & \multicolumn{5}{c}{Dependent variable: Mainnet Transactions per Second} \\
\cmidrule(lr){2-6}
 & Coef. & Std. Err. & $z$ & $p$-value & [0.025, 0.975] \\
\midrule
Constant 
& 2.67 & 1.26 & 2.12 & 0.03 & [0.20,\;5.15] \\
Gas Limit 
& $3.52\times10^{-7}$ & $3.25\times10^{-8}$ & 10.83 & 0.000 & [$2.88\times10^{-7}$,\;$4.16\times10^{-7}$] \\
\midrule
Observation Period & \multicolumn{5}{c}{2025-01-01 through 2026-03-31} \\
Adjusted $R^2$ & \multicolumn{5}{c}{0.657} \\
\bottomrule
\end{tabular}
\begin{tablenotes}[flushleft]
\item
\texttt{Note:} Gas limit is defined as the median daily gas limit per block. Standard errors are heteroskedasticity and autocorrelation robust (HAC) with lag length selected automatically using a Bartlett kernel.
\end{tablenotes}
\end{threeparttable}
\end{table}

Each additional unit of gas limit capacity is associated with roughly $3.5 \times 10^{-7}$ more transactions processed per second. Therefore, a 10,000,000-unit increase in the gas limit is associated with 3.52 more transactions processed every second. Doubling the gas limit to 120 million would result in 21 additional transactions processed every second on top of the baseline. At the current gas limit of approximately 60 million, the model estimates 24 transactions per second. For March 2026 the average Mainnet TPS was 25.59. 

\begin{figure}[H]
	\centering
		\caption{Ethereum Mainnet TPS: Forecast Using Gas Limit Time Trend}
		\label{fig:mainnet_tps_timetrend}
        \begin{minipage}{0.97\linewidth}
        \begin{center}
			\includegraphics[width=0.97\textwidth]{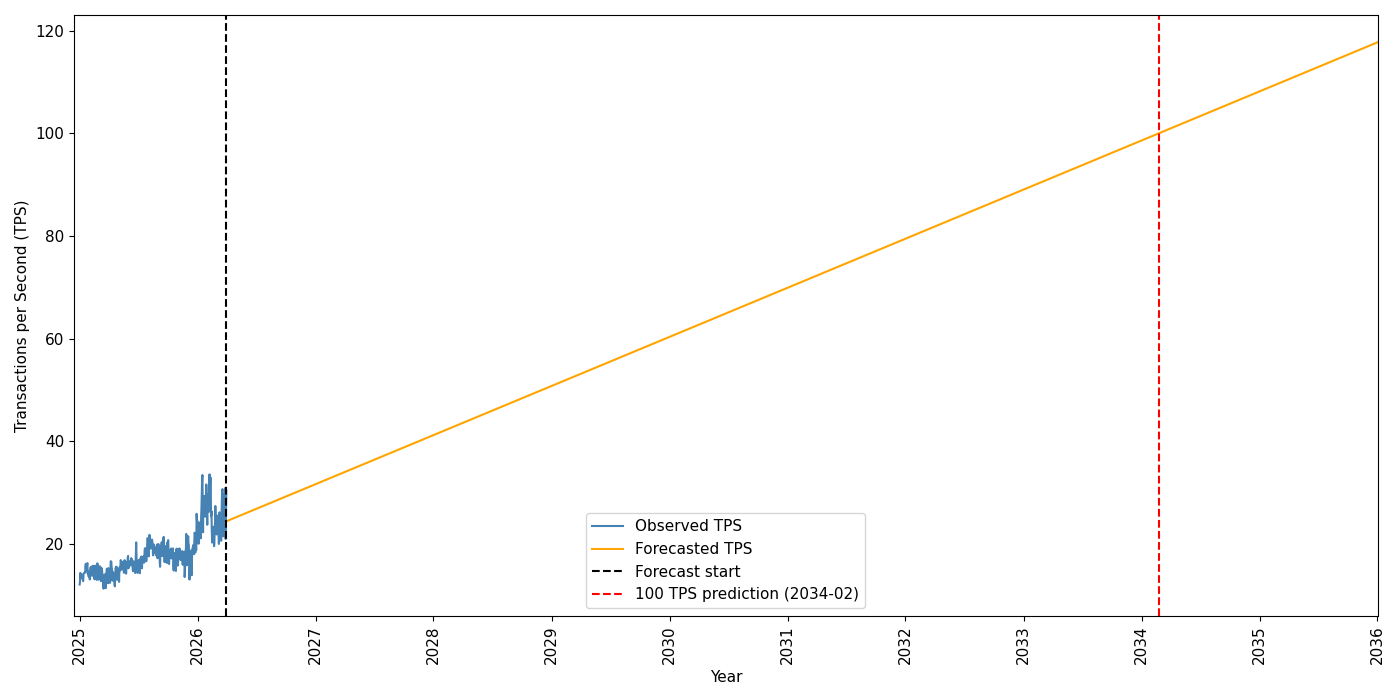} \\
		\end{center}
		\small
		\texttt{Note:}  The blue line denotes the daily TPS on the Mainnet from 2025-01-01 through 2026-03-31. The
        black vertical line represents the beginning of our forecast. The orange line represents our extrapolated TPS using Equation \ref{eq:tps_reduced}. The red vertical line indicates when the predicted value reaches 100 TPS on February 2034. Gas limit data is from our public Dune query: \url{https://dune.com/queries/7347447}. Mainnet TPS is calculated using daily transaction counts from Google BigQuery's public Ethereum transaction data.
    \end{minipage}
\end{figure}

Given the substantial gap between the Mainnet and Solana TPS, we accordingly narrow the analysis to forecast when the Mainnet will reach 100 TPS. Solving for the gas limit required to achieve 100 TPS using Table~\ref{tab:tps_gas} estimated coefficients yields:
\begin{equation}
\text{Gas Limit}^* = \frac{100 - 2.67}{3.52 \times 10^{-7}} \approx 278{,}000{,}000.
\end{equation}
\noindent The model predicts that 100 TPS on the Mainnet requires a gas limit 
of approximately 278 million. At the current gas limit of approximately 60 million, 
this represents a 4.6-fold increase from current levels. Substituting the predicted gas limit values using Table \ref{tab:gas_trend} results, we define our model as follows:
\begin{align}
\text{TPS}_{\text{Mainnet},\, t+\tau}
&= 2.67 + 3.5 \times 10^{-7} \cdot \text{Gas Limit}_{t+\tau} \notag \\
&= 2.67 + 3.5 \times 10^{-7} \cdot \left(2.78 \times 10^{7} + 7.44 \times 10^{4} \cdot (t+\tau)\right) \notag \\
&\approx 12.40 + 0.03 \cdot (t+\tau)
\label{eq:tps_reduced}
\end{align}
where $t = 1, 2, \ldots, T$ with $t=1$ corresponding to 2025-01-01 and $T=454$ corresponding to 2026-03-31, and $\tau \geq 0$ indexing days into the forecast horizon. We predict the Mainnet to reach 100 TPS by February 2034 as seen in Figure \ref{fig:mainnet_tps_timetrend}.

\subsection{Gigagas}
Under the current growth trajectory, 100 TPS will take almost eight years to reach. However, the Ethereum Foundation has outlined a more aggressive path forward with a goal to reach one gigagas per second.

The L1 Strawmap proposes an increase of the gas limit by a factor of ten every two years, totaling a 100-fold increase over four years under EIP-7938.\footnote{\cite{eip7938}} This upgrade is not expected to activate until 2027. Therefore, we use the time trend model in Table \ref{tab:gas_trend} to forecast the gas limit until 2027, at which point the EIP formula is applied.

\begin{equation}
\text{Gas Limit}_{t+\tau} =
\begin{cases}
2.78\times10^{7} + 7.44\times10^{4}\cdot(t+\tau) & \tau \leq 275 \\
\text{Gas Limit}_{t+276} \cdot 10^{(\tau-275)/730} & 276 \leq \tau \leq 1736 \\
\text{Gas Limit}_{t+276}  \cdot 100 & \tau > 1736
\label{eq:3lines}
\end{cases}
\end{equation}

\noindent where $t = 1, 2, \ldots, 454$ with $t=1$ corresponding to 2025-01-01 and $t=454$ corresponding to 2026-03-31, and $\tau \geq 0$ indexing days into the forecast horizon. $\tau=276$ corresponds to 2027-01-01, $\tau=1736$ corresponds to 2030-12-31. The gas limit increases by a factor of ten every two years, with the upper threshold at $100 \times \text{Gas Limit}_{t+276}$ after four years. Mainnet 
TPS is then forecasted using the first line of Equation \ref{eq:tps_reduced}
\begin{equation}
\text{TPS}_{\text{Mainnet},\, t+\tau}
= 2.67 + 3.5 \times 10^{-7} \cdot \text{Gas Limit}_{t+\tau} \notag \\
\end{equation}

\begin{figure}[H]
	\centering
		\caption{Gas Limit: Observed vs Predicted (EIP-7938)}
		\label{fig:mainnet_tps_timetrend_gigagas}
        \begin{minipage}{0.97\linewidth}
        \begin{center}
			\includegraphics[width=0.97\textwidth]{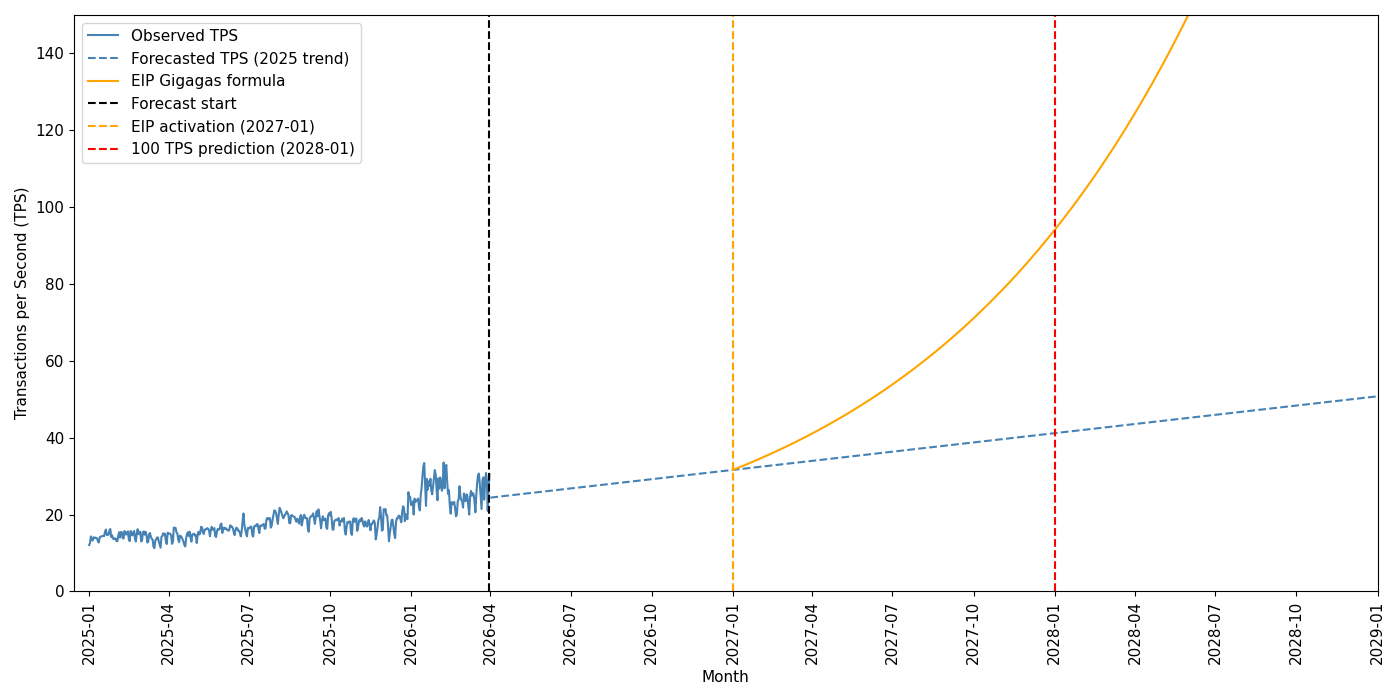} \\
		\end{center}
		\small
		\texttt{Note:} The blue line denotes the daily TPS on the Mainnet from 2025-01-01 through 2026-03-31.  The black vertical line represents the beginning of our forecast. The blue dashed line denotes our extrapolated TPS using our gas limit time trend Equation \ref{eq:gaslim} and TPS model Equation \ref{eq:tps_reduced}. The dashed yellow vertical line at 2027-01-01 is the beginning of the EIP-7938 gas limit increase rule. The solid yellow line denotes our extrapolated TPS using the gas limit increase rule in Equation \ref{eq:3lines}. Gas limit data is from our Dune query: \url{https://dune.com/queries/7347447} and the Mainnet TPS is calculated using daily transaction counts from Google BigQuery's public Ethereum transaction data.
    \end{minipage}

\end{figure}

Figure \ref{fig:mainnet_tps_timetrend_gigagas} presents the forecasted gas limit under the combined 2025-01-01 to 2026-03-31 trendline and EIP-7938 exponential growth schedule. 
The gas limit grows gradually through 2027 following the observed 2025 trend, after which the EIP formula takes effect, producing a sharp exponential increase. Under EIP-7938, the Mainnet reaches 100 TPS by January 2028, more than six years earlier than when using current trends.  The gas limit is projected to level off at 8.22 billion at the end of 2030 with the Mainnet at 2,896 TPS. 

\subsection{Transactions Per Second on the L2 Networks}
\label{sec:Transactions Per Second on the L2 Networks}
\begin{figure}[H]
	\centering
		\caption{Layer 2 Total TPS and Target Blob Counts}
		\label{fig:layer2_tps_target_blob}
        \begin{minipage}{0.97\linewidth}
        \begin{center}
			\includegraphics[width=0.97\textwidth]{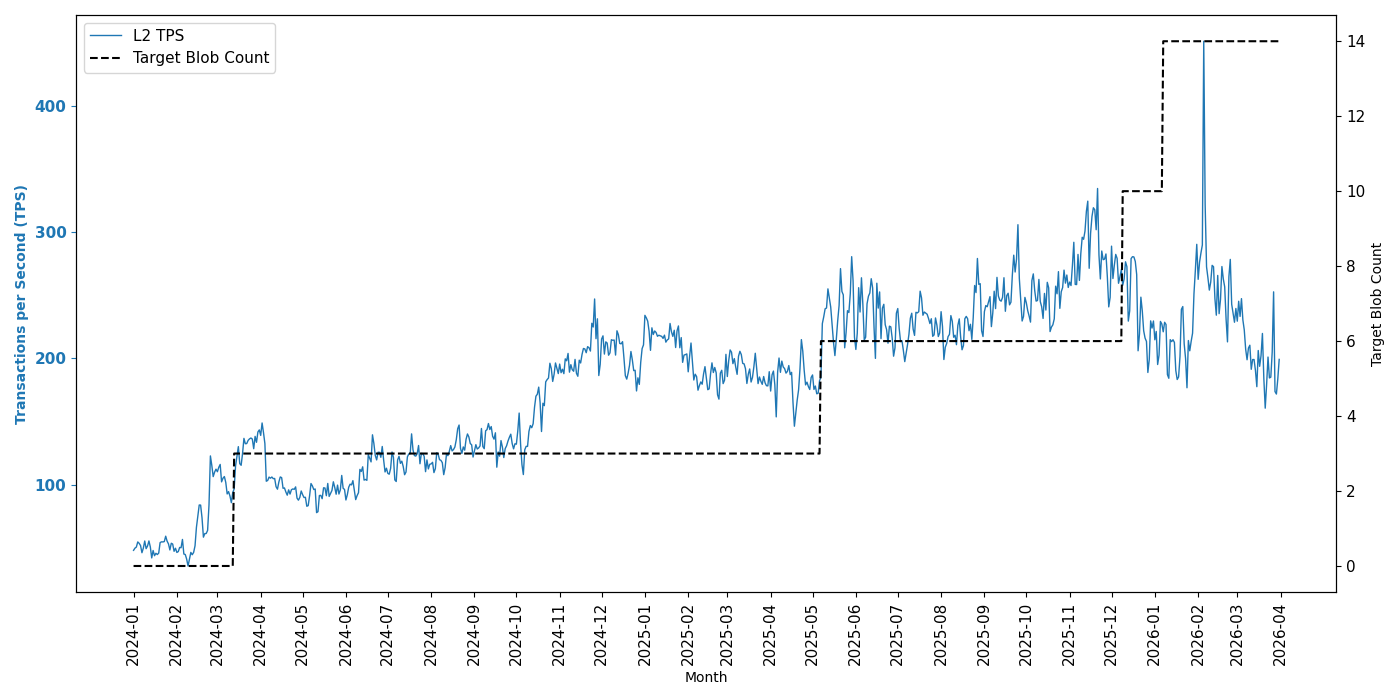} \\
		\end{center}
		\small
		\texttt{Note:} The blue line on the primary y-axis denotes the daily TPS of all the L2 networks. The black dotted line on the secondary y-axis denotes the daily target blob count. TPS data is obtained from our public Dune query: \url{https://dune.com/queries/6962020}, and target blobs are from Figure \ref{fig:blob_target_max_avg}.
    \end{minipage}
\end{figure}

We now study the impact of the target blob count on L2 total TPS.  Figure \ref{fig:layer2_tps_target_blob} plots L2 total TPS alongside the target blob count from January 2024 through March 2026.

 The step increases in the target blob count correspond to the Dencun upgrade in March 2024, the Pectra upgrade in May 2025, and the BPO activations in December 2025 and January 2026. From January 2024 through November 2025, TPS rose broadly in line with each increase in the target blob count. Following the second BPO activation in January 2026, TPS spiked sharply before declining through 2026 Q1, ending with TPS below pre-BPO levels. Table~\ref{tab:tps_blobs} quantifies the impact target blobs have on TPS.

\begin{table}[H]
\centering
\caption{Effect of Target Blobs on L2 TPS}
\label{tab:tps_blobs}
\begin{threeparttable}
\begin{tabular}{lccccc}
\toprule
 & \multicolumn{5}{c}{Dependent variable: Layer 2 Total TPS} \\
\cmidrule(lr){2-6}
 & Coef. & Std. Err. & $z$ & $p$-value & [0.025, 0.975] \\
\midrule
Constant 
& 128.46 & 8.65 & 14.86 & 0.000 & [111.51,\;145.41] \\
Target Blobs 
& 10.75 & 1.52 & 7.09 & 0.000 & [7.78,\;13.72] \\
\midrule
Observation Period & \multicolumn{5}{c}{2024-01-01 through 2026-03-31} \\
Adjusted $R^2$ & \multicolumn{5}{c}{0.37} \\
\bottomrule
\end{tabular}
\begin{tablenotes}[flushleft]
\item
\texttt{Note:} Standard errors are heteroskedasticity and autocorrelation robust (HAC) with lag length selected automatically by a Bartlett kernel.
\end{tablenotes}
\end{threeparttable}
\end{table}

\noindent Each additional target blob is associated with 10.75 more L2 total TPS. Increasing the blob count by four, as it was with both BPOs, results in an increase of 43 TPS. At the current target blob count of 14, the model predicts a baseline of 278.96 L2 total TPS. The average TPS since the second BPO activation was 227.77.

We now turn to our forecast of TPS and construct our projection to maintain the stepwise nature of the target blob increases. BPO-1 and BPO-2 each added four blobs to the target, 30 days apart, giving an increment rate of four blobs per 30 days defined as:

\begin{equation}
 \label{eq:blob}
     \text{Target Blobs}_{t+ \tau} = \min\left(14 + int\left\lfloor \frac{t + \tau}{30} \right\rfloor \times 4,\ 128\right)
\end{equation}
such that $t = 1, 2, \ldots, T$ with $t=1$ corresponding to 2024-01-01 and $T=820$ corresponding to 2026-03-31, and $\tau \geq 0$ indexes days into the forecast horizon. We apply this same increment schedule going forward until the target blob count reaches 128.\footnote{The Strawmap projects one gigagas per second for the Layer 2s, so this target is very conservative} The projected target blob count is then used to forecast L2 TPS using the coefficients from Table~\ref{tab:tps_blobs}:
\begin{equation}
\label{eq:expblob}
\text{TPS}_{\text{L2},t+\tau} = 128.46 + 10.75 \cdot \text{TargetBlobs}_{t+\tau}
\end{equation}
where $\text{TPS}_{\text{L2},t+\tau}$ denotes L2 total TPS and $\text{TargetBlobs}_{t+\tau}$ denotes the projected target blob count from Equation \ref{eq:blob}.

\begin{figure}[H]
	\centering
		\caption{TPS L2 Total: Observed vs Predicted}
		\label{fig:layer2_tps_prediction}
        \begin{minipage}{0.97\linewidth}
        \begin{center}
			\includegraphics[width=0.97\textwidth]{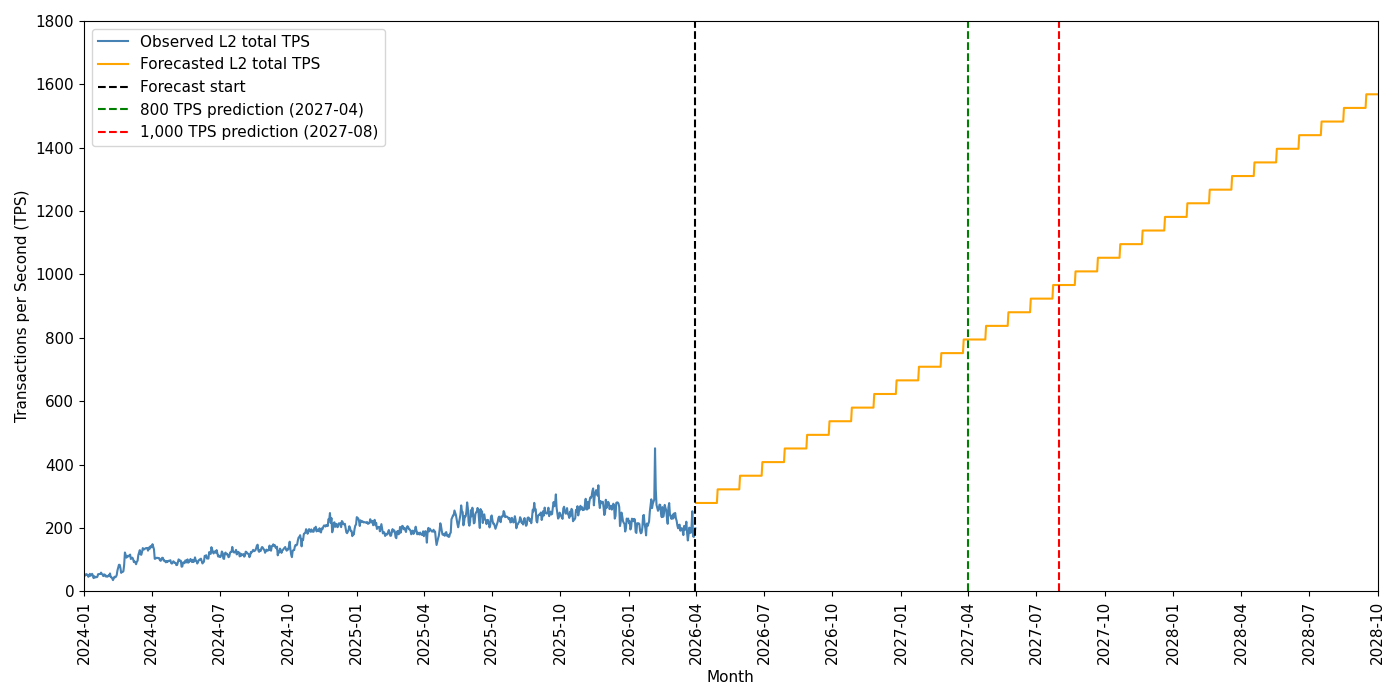} \\
		\end{center}
		\small
		\texttt{Note:} The blue line denotes the daily total TPS across all L2 networks from 2024-01-01 through 2026-03-31. The black vertical line represents the beginning of our forecast. The orange line represents the L2 total forecasted TPS using Equation \ref{eq:expblob}. The green vertical line indicates when the predicted value reaches 800 TPS in April 2027. The red vertical line indicates when the predicted value reaches 1,000 TPS in August 2027. Daily data for the individual L2 networks is obtained from our public Dune query: \url{https://dune.com/queries/6962020}.
    \end{minipage}
\end{figure}

Figure \ref{fig:layer2_tps_prediction} displays the forecasted L2 total TPS under our defined blob count increase structure.

L2 total is expected to reach 800 TPS in April 2027 at 66 blobs and 1,000 TPS on August 2027 at 82 blobs. 

\subsection{Solana Crossover Prediction}

We now examine when L2 total will close the speed gap with Solana. For 2026 Q1, Solana had an average daily TPS of 1,303.46 and the combined total for the L2s was 226.92, a difference of 1,076.54 TPS.

The L2 total forecast is directly from Section \ref{sec:Transactions Per Second on the L2 Networks}. Solana TPS is extrapolated using a linear time trend fit to historical data from January 1, 2024 to March 31, 2026: 
\begin{equation}
\text{TPS}_{\text{Solana},t+\tau} = 745.92 + 0.56 \cdot (t+\tau)
\label{eq:solforecast}
\end{equation}
where $t = 1, 2, \ldots, T$ with $t=1$ corresponding to 2024-01-01 and $T=820$ corresponding 
to 2026-03-31, and $\tau \geq 0$ indexes days into the forecast horizon.

\begin{figure}[H]
	\centering
		\caption{L2 Total and Solana TPS Prediction}
		\label{fig:layer2_solana_tps_crossover}
        \begin{minipage}{0.97\linewidth}
        \begin{center}
			\includegraphics[width=0.97\textwidth]{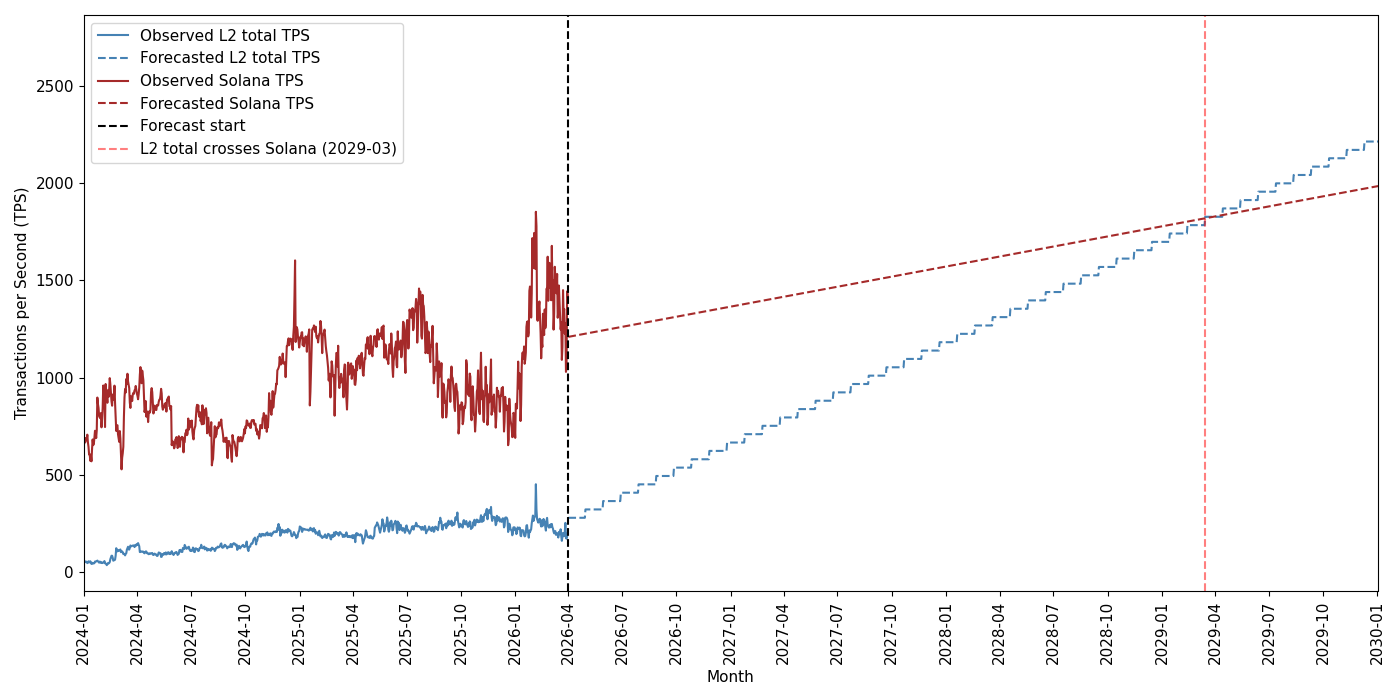} \\
		\end{center}
		\small
		\texttt{Note:} The brown line denotes the daily TPS on Solana and the blue line denotes the daily TPS of L2 total from 2024-01-01 through 2026-03-31. The black vertical line represents the beginning of our forecast. The dashed brown and blue lines represent the forecasted TPS for Solana using Equation \ref{eq:solforecast} and L2 total using Equation  \ref{eq:expblob}, respectively. The red vertical line indicates when Solana and L2 total cross in March 2029. Daily data for Solana is obtained from our public Dune query: \url{https://dune.com/queries/6946173}. Daily data for the individual L2 total networks is obtained from our public Dune query: \url{https://dune.com/queries/6962020}. 

    \end{minipage}
\end{figure}

Figure \ref{fig:layer2_solana_tps_crossover} presents historical and forecasted TPS for L2 Total and Solana until 2030.

 L2 total TPS is projected to surpass Solana TPS in March 2029 with both networks at approximately 1,820 TPS at the crossing point.

\section{Median Transaction Fee Forecast}
\label{sec:feeforecast}
The Mainnet is not expected to reach Solana speeds within the next 10 years under the current gas limit trajectory, but the outlook for transaction fees is more favorable. This section will cover predictions for Mainnet fees and compare them with Solana's forecasted fees.

\begin{figure}[H]
	\centering
		\caption{Ethereum Median Transaction Fee and Block Gas Limit}
		\label{fig:mainnet_fees_gas_limits}
        \begin{minipage}{0.97\linewidth}
        \begin{center}
			\includegraphics[width=0.97\textwidth]{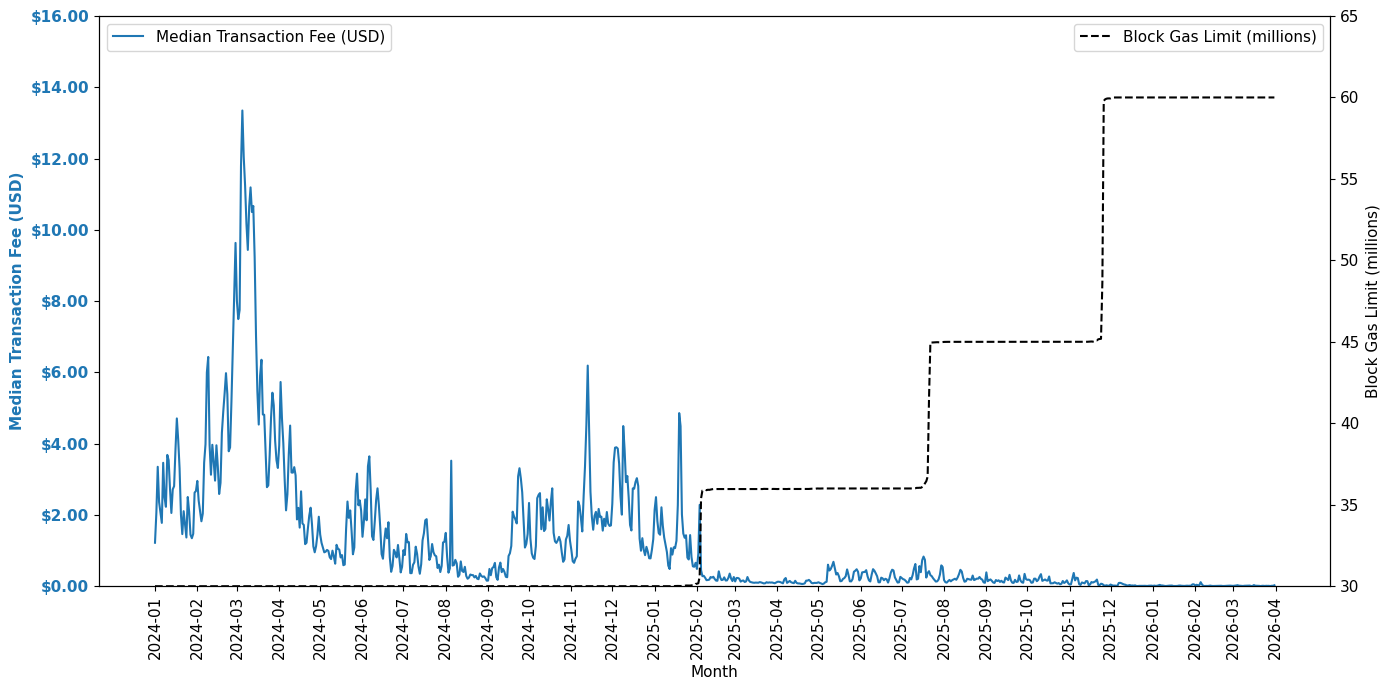} \\
		\end{center}
		\small
		\texttt{Note:} The median daily transaction fee in USD is plotted in blue on the primary y-axis. The black dashed line is the block gas limit which begins at 30 million units on the secondary y-axis. Mainnet transaction fees in ETH are collected from Google BigQuery public Ethereum transaction data. The daily closing price of ETH is then collected from CoinGecko to convert fees to USD. Gas limit data is collected using our Dune query: \url{https://dune.com/queries/7347447}.
    \end{minipage}
\end{figure}

Figure \ref{fig:mainnet_fees_gas_limits} plots the daily median transaction fees on the Mainnet from January 1, 2024, through March 31, 2026.

\subsection{Ethereum Mainnet}
\label{sec:EM}

 Fees have declined substantially over this period. The daily median transaction fee for 2024 Q1 was \$3.786300 and the fee for 2026 Q1 was \$0.012283, a 99.68\% decline. Overall, as the gas limit increased, median transaction fees declined. Between 2024-01-01 and 2025-02-03, the gas limit was roughly 30 million. The median transaction fee for this period was \$1.317987. Between 2025-02-04 and 2025-07-20 the gas limit was roughly 36 million. The median transaction fee for this period was 
 \$0.172602. Between 2025-07-21 and 2025-11-25 the gas limit was roughly 45 million. The median transaction fee for this period was \$0.162836. Between 2025-11-26 and 2026-03-31 the gas limit was roughly 60 million. The median transaction fee for this period was \$0.012753.
 
As of 2026 Q1, the median transaction fee is \$0.012283. We are interested in forecasting when the Mainnet median fee will fall to \$0.0005, which is the 2026 Q1 Solana median fee. Estimation results are in Table \ref{tab:log_fee_reg}, which reports the impact of the gas limit on the log median transaction fee. We take the log of the median fee to ensure that forecasted values remain positive.

\begin{table}[H]
\centering
\caption{Determinants of Ethereum Median Transaction Fees (USD)}
\label{tab:log_fee_reg}
\begin{threeparttable}
\begin{tabular}{lccccc}
\toprule
 & \multicolumn{5}{c}{Dependent variable: $\log(\text{Median Fee USD})$} \\
\cmidrule(lr){2-6}
 & Coef. & Std. Err. & $z$ & $p$-value & [0.025, 0.975] \\
\midrule
Constant 
& 3.01 & 0.32 & 9.37 & 0.000 & [2.38,\;3.64] \\
$\text{Gas Limit}$ 
& $-1.19\times10^{-7}$ & $6.54\times10^{-9}$ & -18.18 & 0.000 & [$-1.32\times10^{-7}$,\;$-1.06\times10^{-7}$] \\
\midrule
Observation Period & \multicolumn{5}{c}{01-01-2025 through 03-31-2026} \\
Adjusted $R^2$ & \multicolumn{5}{c}{0.74} \\
\bottomrule
\end{tabular}
\begin{tablenotes}[flushleft]
\item
\texttt{Note:} Gas limit is defined as the median daily gas limit per block. Standard errors are heteroskedasticity and autocorrelation consistent (HAC) with lag length selected automatically by a Bartlett kernel.
\end{tablenotes}
\end{threeparttable}
\end{table}

A 5-million-unit increase in the gas limit is associated with fees falling by approximately 59.2\%.  At a gas limit of 60 million, the model predicts that the median fee will be \$0.016, which is quite close to the Mainnet's average median fee of \$0.013 since the gas limit reached 60 million on 2025-11-26.

To forecast the median fee, 
we extrapolate the gas limit from the 2025 time trend as estimated in Table 
\ref{tab:gas_trend}. Therefore, recall the gas limit forecast model: 
\begin{equation}
\label{eq:blob2}
\text{Gas Limit}_{t+\tau} = 2.78\times10^{7} + 7.44\times10^{4}\cdot(t+\tau)
\end{equation}

\noindent where $t = 1, 2, \ldots, T$ with $t=1$ corresponding to 2025-01-01 and $T=454$ corresponding 
to 2026-03-31, and $\tau \geq 0$ indexes days into the forecast horizon. Our fee forecast model then becomes:
\begin{align}
\ln(\text{Fee}^{\text{Mainnet}}_{t+\tau})
&= 3.01 - 1.19\times10^{-7} \cdot \text{GasLimit}_{t+\tau} \notag \\
&= 3.01 - 1.19\times10^{-7} \cdot \left(2.78\times10^{7} + 7.44\times10^{4}\cdot(t+\tau)\right) \notag \\
&\approx -0.30 - 0.00885\cdot(t+\tau).
\label{eq:fee_reduced}
\end{align}
$\text{GasLimit}_{t+\tau}$ is given by Equation \ref{eq:blob2} and $\text{Fee}^{\text{Mainnet}}_{t+\tau}$ denotes the daily median transaction fee of the Mainnet.

\begin{figure}[H]
	\centering
		\caption{Median Transaction Fee (USD): Observed vs Predicted}
		\label{fig:mainnet_fee_prediction}
        \begin{minipage}{0.97\linewidth}
        \begin{center}
			\includegraphics[width=0.97\textwidth]{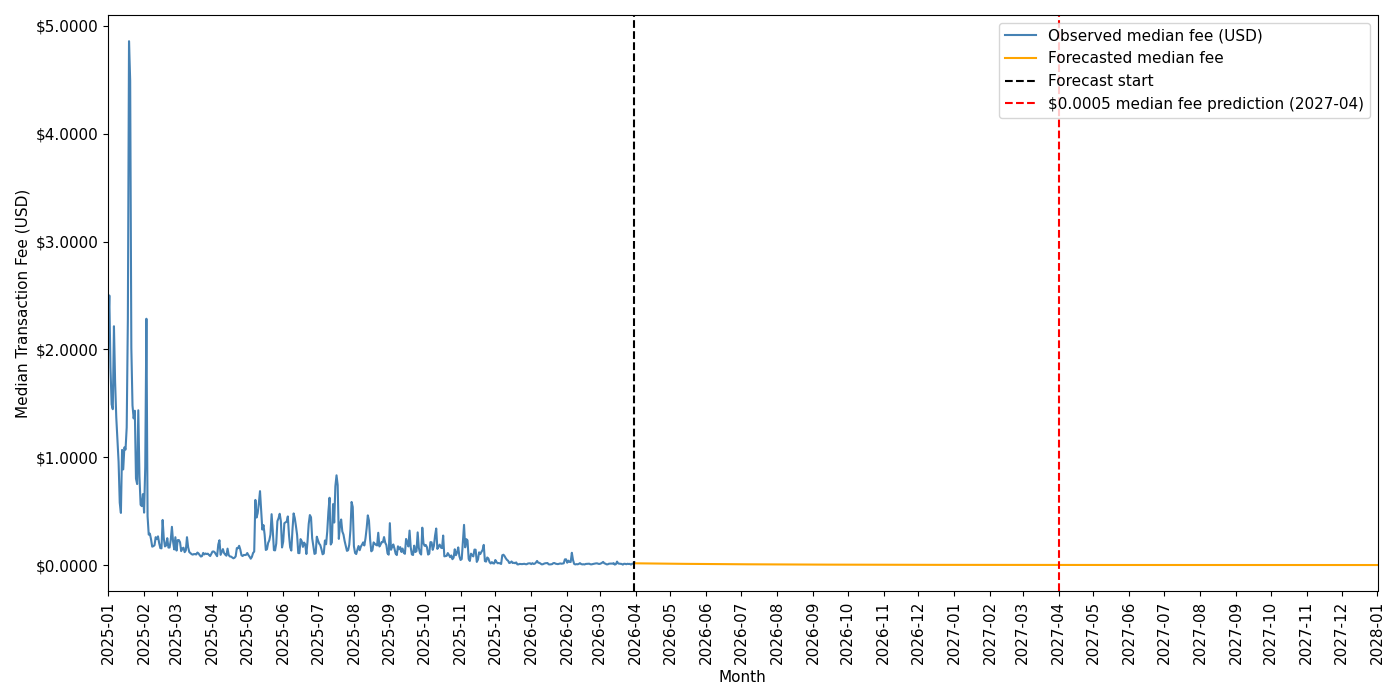} \\
		\end{center}
		\small
		\texttt{Note:} The blue line denotes the daily median transaction fee in USD on the Mainnet from 2025-01-01 through 2026-03-31. The black vertical line represents the beginning of our forecast. The orange line represents the forecasted median fees for the Mainnet using Equation \ref{eq:fee_reduced}. The red dotted line indicates when the predicted value reaches \$0.0005 in April 2027. Daily data for the Mainnet median transaction fee in ETH is obtained from Google BigQuery’s public Ethereum transaction data. We convert fees in ETH to USD using daily closing prices from CoinGecko.
    \end{minipage}
\end{figure}

Figure \ref{fig:mainnet_fee_prediction} displays the forecasted Mainnet median transaction fees under our model Equation \ref{eq:fee_reduced} for the effect of gas limits on fees.

\noindent Mainnet median fees are expected to reach \$0.0005 in April 2027.

\subsection{Ethereum and Solana Cross Over Prediction}
Having shown that Mainnet fees are projected to reach Solana's 2025 median fee level by early 2027, we now examine when Mainnet fees can converge with Solana's projected fee level.  

Our Mainnet median fee forecast is directly from Section \ref{sec:EM}. We forecast Solana fees by fitting a time trend to the log of historical fees. Our model is as follows:
\begin{equation}
\label{eq:feeSolana}
\ln(\text{Fee}^{\text{Solana}}_{t+\tau}) = -6.49 - 0.0021\cdot(t+\tau)
\end{equation}
where $t = 1, 2, \ldots, T$ with $t=1$ corresponding to 2025-01-01 and $T=454$ corresponding 
to 2026-03-31, and $\tau \geq 0$ indexes days into the forecast horizon. $\ln(\text{Fee}^{\text{Solana}}_{t+\tau})$ denotes the logged daily median transaction fee on Solana.

\begin{figure}[H]
	\centering
		\caption{Ethereum and Solana Crossover Prediction}
		\label{fig:eth_sol_fee_crossover}
        \begin{minipage}{0.97\linewidth}
        \begin{center}
			\includegraphics[width=0.97\textwidth]{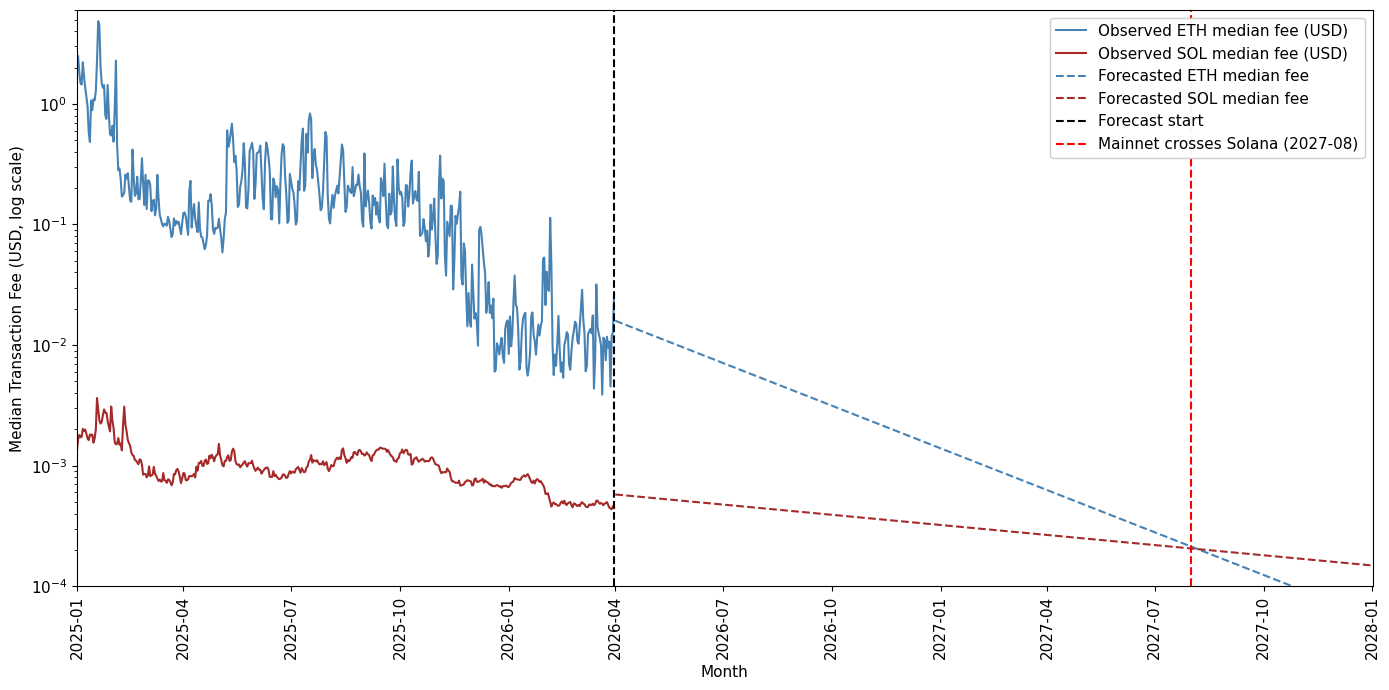} \\
		\end{center}
		\small
		\texttt{Note:} The brown line denotes the daily median transaction fee in USD on Solana and the blue line denotes the daily median transaction fee in USD on the Mainnet from 2025-01-01 through 2026-03-31. Fees are transformed to log values for visualization. The black vertical line indicates the beginning of our forecast. The dashed brown and blue lines represent the forecasted median fees for Solana using Equation \ref{eq:feeSolana} and the Mainnet using Equation \ref{eq:fee_reduced}, respectively. The red vertical line indicates when Solana and the Mainnet cross in August 2027. Daily fee data in USD for Solana is obtained from our public Dune query: \url{https://dune.com/queries/6946173}. Daily data for the Mainnet median transaction fee in ETH is obtained from Google BigQuery’s public Ethereum transaction data. We convert fees in ETH to USD using daily closing prices from CoinGecko.

    \end{minipage}
\end{figure}

Figure \ref{fig:eth_sol_fee_crossover} presents historical and forecasted daily median transaction fees for the Mainnet and Solana until 2028. 
The Mainnet is projected to achieve Solana median transaction fees in August 2027, with both networks' fees at \$0.000204.

\subsection{Layer 2 Leader Fee Trends and Solana Crossover}
Fees on the cheapest of the largest L2 networks are already below Solana levels in the first quarter of 2026. Optimism, the third largest L2, has fees of \$0.000027 for 2026 Q1, 94.6\% lower than Solana’s at \$0.000496.

We consider now when the larger set of L2 leaders, adding Base and Arbitrum to Optimism, will reach fee Solana fee levels.  We use the same explanatory variables that we use for L2 TPS, regressing log fees on target blobs in Table \ref{tab:fee_blobs}.     

\begin{table}[H]
\centering
\caption{Effect of Target Blobs on L2 Fees}
\label{tab:fee_blobs}
\begin{threeparttable}
\begin{tabular}{lccccc}
\toprule
 & \multicolumn{5}{c}{Dependent variable: Layer 2 Leader Fees} \\
\cmidrule(lr){2-6}
 & Coef. & Std. Err. & $z$ & $p$-value & [0.025, 0.975] \\
\midrule
Constant 
& -6.0220  &    0.118   &  -51.104   & 0.000      &[-6.253,\;-5.791] \\
Target Blobs 
& -0.0476    &  0.013    & -3.803    &  0.000      & [-0.072,\;-0.023] \\
\midrule
Observation Period & \multicolumn{5}{c}{2025-01-01 through 2026-03-31} \\
Adjusted $R^2$ & \multicolumn{5}{c}{0.09} \\
\bottomrule
\end{tabular}
\begin{tablenotes}[flushleft]
\item
\texttt{Note:} Standard errors are heteroskedasticity and autocorrelation robust (HAC) with lag length selected automatically by a Bartlett kernel.
\end{tablenotes}
\end{threeparttable}
\end{table}

\noindent Our estimates in Table \ref{tab:fee_blobs} imply that every 30 days when the blob target rises by four, L2 median fees fall -17.34\%.

We create a forecast using the coefficients from Table \ref{tab:fee_blobs} and our blob growth rule

\begin{equation}
\ln(\text{Fee}^{\text{L2}}_{t+\tau}) = -6.0220 - 0.00476 \cdot \text{TargetBlobs}_{t+\tau}
\label{eq:l2leaderf}
\end{equation}

\noindent where $\text{Fee}^{\text{L2}}_{t+\tau}$ denotes L2 leader fees and $\text{TargetBlobs}_{t+\tau}$ denotes the projected target blob count from Equation \ref{eq:blob}.

\begin{figure}[H]
	\centering
		\caption{L2 Leader Fee Forecast}
		\label{fig:layer2_fees_blob_target_forecast}
        \begin{minipage}{0.97\linewidth}
        \begin{center}
			\includegraphics[width=0.97\textwidth]{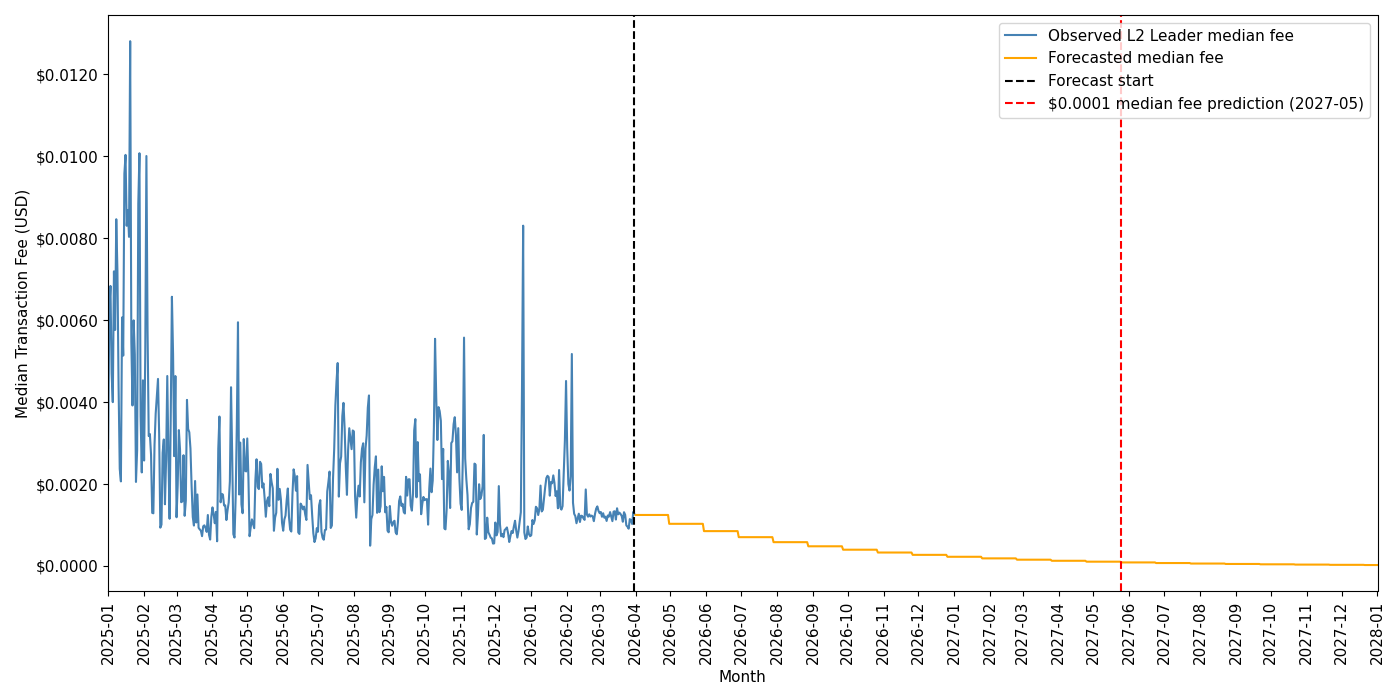} \\
		\end{center}
		\small
		\texttt{Note:} The blue line denotes the daily median transaction fee in USD of the L2 leader average from 2025-01-01 through 2026-03-31. The black vertical line represents the beginning of our forecast. The orange line represents the forecasted median fees for the L2 leaders using Equation \ref{eq:l2leaderf}. The red vertical line indicates when the predicted value reaches \$0.0001 in May 2027. Daily data for the individual L2 leaders is obtained from our public Dune query: \url{https://dune.com/queries/6962020}.

    \end{minipage}
\end{figure}

Figure  \ref{fig:layer2_fees_blob_target_forecast} shows that the leading L2 networks will reach \$0.0001 fees in May 2027.

\begin{figure}[H]
	\centering
		\caption{L2 Leader and Solana Crossover Prediction}
		\label{fig:layer2_solana_fee_crossover}
        \begin{minipage}{0.97\linewidth}
        \begin{center}
			\includegraphics[width=0.97\textwidth]{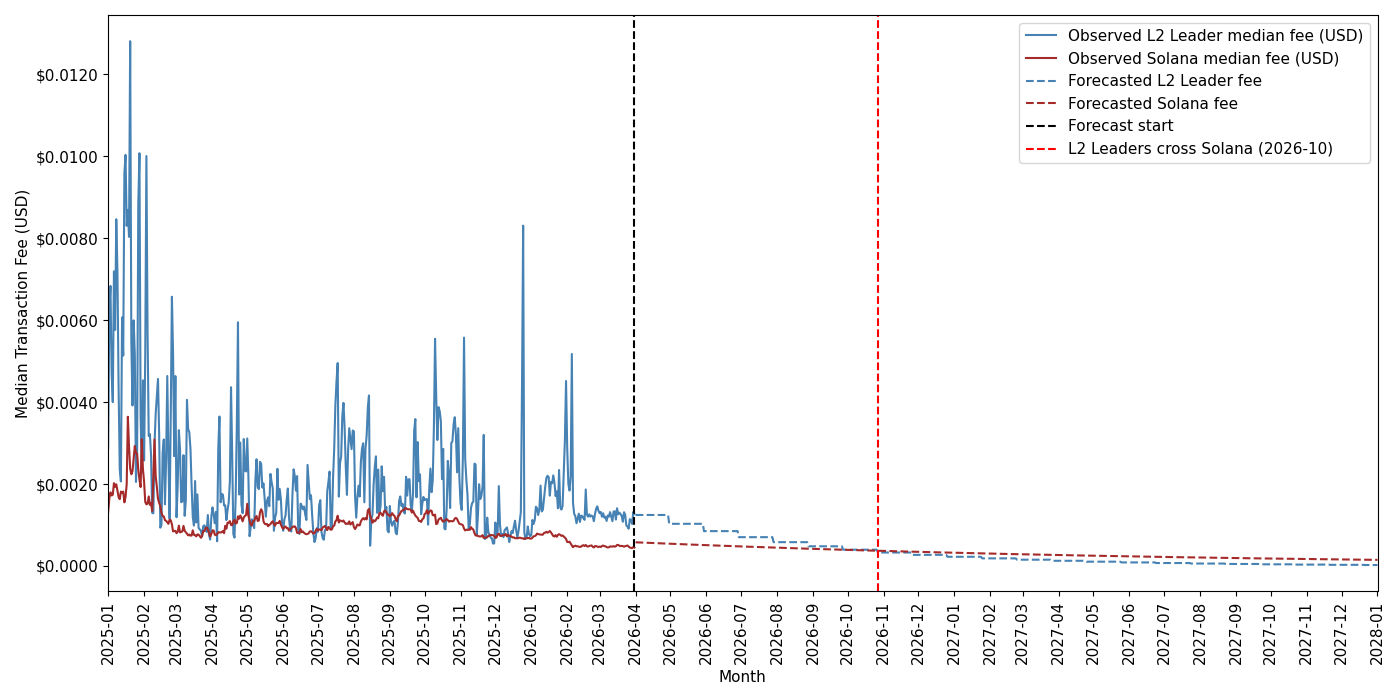} \\
		\end{center}
		\small
		\texttt{Note:} The brown line denotes Solana's daily median transaction fee in USD, and the blue line denotes the daily median transaction fee in USD of the L2 leader average from 2025-01-01 through 2026-03-31.  The black vertical line represents the beginning of our forecast. The dashed brown and blue lines represent the forecasted median fees for Solana using Equation \ref{eq:feeSolana} and the L2 leader average using Equation \ref{eq:l2leaderf}, respectively. The red vertical line indicates when Solana and the L2 leader average cross in October 2026. Daily data for Solana is obtained from our public Dune query: \url{https://dune.com/queries/6946173}. Daily data for the individual L2 leaders is obtained from our public Dune query: \url{https://dune.com/queries/6962020}.

    \end{minipage}
\end{figure}

In predicting a crossover with Solana, we use the same time trend model Equation \ref{eq:feeSolana}. Figure \ref{fig:layer2_solana_fee_crossover}
displays our predicted crossover in October 2026.

Layer 2 fees are falling more quickly than Solana's. Our forecast predicts that the Layer 2 per transaction fees of \$0.000328 cross Solana's at 42 blobs in October 2026.

\section{Conclusion}
\label{sec:con}
This paper documents the evolution of transaction speed and cost from January 2024 through March 2026 across the Ethereum network, its L2 ecosystem, Polygon, and Solana. This period is defined by Dencun, Fusaka, Pectra, three major Ethereum protocol upgrades, and BPO-1 and BPO-2, two blob parameter-only upgrades. 

In terms of speed, Solana remains the fastest network, averaging 1,303 TPS in 2026 Q1. Mainnet speed has nearly doubled since 2024, reaching 25.78 TPS in 2026 Q1. We estimate that each 10-million-unit increase in the gas limit is associated with 3.52 additional TPS, meaning the Mainnet cannot reach 100 TPS under current gas limit trajectories until February 2034. The proposed Strawmaps's EIP-7938 gas limit increase, if activated in 2027, would accelerate that timeline considerably, allowing the Mainnet to reach 100 TPS by January 2028.

We anticipate a much faster growth in TPS for the L2 networks. Extending the BPO upgrade schedule, with a 4-blob increase every 30 days, L2 TPS is projected to cross Solana's time-trend extrapolated TPS by March 2029 at 1,820 TPS.

Our analysis indicates the Mainnet should reach Solana fee levels in less than two years. A 5-million-unit increase in the gas limit is associated with fees falling by approximately
59.2\%. Even if EIP-7938 is not adopted, the Mainnet will reach Solana's extrapolated time-trend fees of \$0.0002 by August 2027. The L2 networks are currently competitive with Solana on fees, but we forecast declining fees across the leading networks as target blobs increase.  We project that average fees across Arbitrum, Base and Optimism to fall below Solana's by October 2026.

With projected gas limit increases, the Mainnet may soon be competitive with Solana on fees. Without a healthy L2 ecosystem though, the Mainnet will be a sluggish alternative to faster blockchains. It will likely continue to lag Solana on speed-sensitive DEX trading.

\newpage
\bibliography{resources}
\end{document}